\pdfoutput=1

\documentclass[aps, prd, reprint, twocolumn, groupedaddress, superscriptaddress, showpacs, nofootinbib]{revtex4-1}

\usepackage{lmodern}
\usepackage{amsmath, amssymb}
\usepackage{slashed}
\usepackage{graphicx}
\usepackage{caption}
\usepackage{subcaption}
\usepackage{hyperref}
\usepackage{multirow}
\usepackage{epstopdf}

\newcommand{\sect}[1]{\section{#1}}

\begin{document}

\title{NLO Effects for Doubly Heavy Baryon in QCD Sum Rules}

\date{\today}

\author{Chen-Yu \surname{Wang}}
\affiliation{School of Physics and State Key Laboratory of Nuclear Physics and Technology, Peking University, Beijing 100871, China}
\author{Ce \surname{Meng}}
\affiliation{School of Physics and State Key Laboratory of Nuclear Physics and Technology, Peking University, Beijing 100871, China}
\author{Yan-Qing \surname{Ma}}
\affiliation{School of Physics and State Key Laboratory of Nuclear Physics and Technology, Peking University, Beijing 100871, China}
\affiliation{Center for High Energy Physics, Peking University, Beijing 100871, China}
\affiliation{Collaborative Innovation Center of Quantum Matter, Beijing 100871, China}
\author{Kuang-Ta \surname{Chao}}
\affiliation{School of Physics and State Key Laboratory of Nuclear Physics and Technology, Peking University, Beijing 100871, China}
\affiliation{Center for High Energy Physics, Peking University, Beijing 100871, China}
\affiliation{Collaborative Innovation Center of Quantum Matter, Beijing 100871, China}

\begin{abstract}
    With the QCD sum rules approach, we study the newly discovered doubly heavy baryon $\Xi_{cc}^{++}$.
    We analytically calculate the next-to-leading order (NLO) contribution to the perturbative part of $J^{P} = \frac{1}{2}^{+}$ baryon current with two identical heavy quarks,
    and then reanalyze the mass of $\Xi_{cc}^{++}$ at the NLO level.
    We find that the NLO correction significantly improves both scheme dependence and scale dependence, whereas it is hard to control these theoretical uncertainties at leading order.
    With the NLO contribution, the baryon mass is estimated to be $m_{\Xi_{cc}^{++}} = 3.66_{-0.10}^{+0.08} \text{~GeV}$, which is consistent with the LHCb measurement.
\end{abstract}


\maketitle

\sect{Introduction}
The quark model predicts rich structures of hadronic states with various flavors.
Numerous predicted states have been observed experimentally, indicating the validity of the quark model classification for hadrons.
However, a class of states, which contain more than one heavy quark, have not been discovered for decades.
Recently, LHCb collaboration observed a highly significant structure in the $\Lambda_{c}^{+} K^{-} \pi^{+} \pi^{+}$ mass spectrum,
which is interpreted as the doubly charmed baryon $\Xi_{cc}^{++}$ \cite{Aaij:2017ueg} with mass $3621 \pm 0.72 \pm 0.27 \pm 0.14 \text{~MeV}$.
Early experimental studies of $\Xi_{cc}^{+}$ were performed by SELEX \cite{Mattson:2002vu}, Babar \cite{Aubert:2006qw}, and Belle \cite{Chistov:2006zj} collaborations.

The understanding of $\Xi_{cc}^{++}$ demands more rigorous theoretical studies.
Plenty of methods have been used in the literature \cite{Hudspith:2017bbh, Namekawa:2013vu, Lewis:2001iz, Sun:2016wzh, Kiselev:2017eic, Shah:2017liu, Gadaria:2016omw, Roberts:2007ni, Ebert:2002ig}.
Among them, the QCD sum rules,
which are based on the first principle of QCD, are powerful tools to study various properties of hadronic states \cite{Shifman:1978bx, Shifman:1978by}.
Many works have been devoted to the study of doubly heavy baryons within QCD sum rules \cite{Bagan:1992za, Kiselev:1999zj, Zhang:2008rt, Wang:2010hs, Tang:2011fv, Aliev:2012ru, Chen:2017sbg},
and some impressive predictions are obtained.
But in all these works, only leading order (LO) in the $\alpha_{s}$ expansion of perturbative contribution and Wilson coefficients of vacuum condensates are considered.
Without higher order contributions, it is hard to control theoretical uncertainties in QCD sum rules, which limits the predictive power.
For instance, at LO, the value of charm quark mass can not be well determined, which will cause large errors.
In fact, it was known a long time ago that the next-to-leading order (NLO) correction has sizable contributions to meson and nucleon sum rules \cite{Reinders:1984sr, Jamin:1987gq, Ovchinnikov:1991mu}.
Therefore, the study of NLO effect for doubly heavy baryons in QCD sum rules is badly needed.

Higher order calculations in QCD sum rules become harder and harder when more particles or more massive particles are involved.
For mesons, the state-of-the-art calculation has been developed to $\mathcal{O}(\alpha_{s}^{4})$ with the help of mass expansion \cite{Schwinger:1989ix, Maier:2011jd, Baikov:2009uw, Chetyrkin:2000zk, Baikov:2008jh, Baikov:2004ku}.
While for baryons, the $\mathcal{O}(\alpha_{s})$ correction is available in the literature only for nucleons and singly heavy baryons \cite{Jamin:1987gq, Ovchinnikov:1991mu, Groote:2008dx}.

In this paper, we calculate the NLO correction to perturbative contribution for the doubly heavy $J^{P} = \frac{1}{2}^{+}$ baryon, and show its important effects in QCD sum rules.
With the help of integration-by-parts method \cite{Chetyrkin:1981qh, Laporta:2001dd} and differential equation method \cite{Henn:2013pwa, Henn:2014qga}, we get a fully analytical expression.
We reproduce the massless result in the literature when we set all quark masses to zero.
Based on this calculation, we reanalyze the newly discovered $\Xi_{cc}^{++}$ in QCD sum rules.

\sect{QCD Sum Rules}
The central object in QCD sum rules is the following two-point correlation function \cite{Shifman:1978bx, Ioffe:1981kw}
\begin{align}
    \Pi(q)
    & =
    i
    \int \mathrm{d}^{4} x \,
    e^{i q x}
    \langle \Omega | T \{ \eta(x) \overline{\eta}(0) \} | \Omega \rangle
    \nonumber \\
    & =
    \Pi_{1}(q^{2}) \slashed{q} + \Pi_{2}(q^{2})
    \, ,
\end{align}
where $\Omega$ denotes the QCD vacuum,
and $\eta$ is the baryon current to be defined later.

On the one hand, one can calculate $\Pi(q)$ using operator product expansion,
which gives
\begin{equation}
    \Pi(q)
    =
    C_{1}(q)
    +
    \sum_{i}
    C_{i}(q) \langle O_{i} \rangle
    \, ,
\end{equation}
where $C_{1}$ is the perturbative contribution and $C_{i}$ is the Wilson coefficient of a gauge invariant Lorentz scalar operator $O_{i}$.
Both $C_{1}$ and $C_{i}$ are perturbatively calculable.
$\langle O_{i} \rangle$ is a shorthand for the vacuum condensates $\langle \Omega | O_{i} | \Omega \rangle$,
which is a nonperturbative but universal quantity.
It means that the value of $\langle O_{i} \rangle$ determined from other processes should be the same as its value in the process considered in this paper.

On the other hand, $\Pi(q)$ satisfies the dispersion relation
\begin{align}
    \Pi(q)
    & =
    \frac{1}{\pi}
    \int_{0}^{\infty} \mathrm{d} s \,
    \frac{\Im \Pi_{1}(s + i \epsilon) \slashed{q} + \Im \Pi_{2}(s + i \epsilon)}{s - q^{2}}
    \nonumber \\
    & =
    \int_{0}^{\infty} \mathrm{d} s \,
    \frac{\rho_{1}(s + i \epsilon) \slashed{q} + \rho_{2}(s + i \epsilon)}{s - q^{2}}
    \, ,
\end{align}
where $\rho_{1}$ and $\rho_{2}$ are the spectrum densities.
Based on the optical theorem, one assumes the spectrum density $\rho(q) = \rho_{1}(q^{2}) \slashed{q} + \rho_{2}(q^{2})$ to be \cite{Ioffe:1981kw}
\begin{equation}
    \rho(q)
    =
    \lambda_{H}^{2}
    (\slashed{q} + m_{H})
    \delta(q^{2} - m_{H}^{2})
    +
    \rho_{c}(q)
    \theta(q^{2} - s_{th})
    \, ,
\end{equation}
where $s_{th}$ is the threshold of continuum spectrum,
$\lambda_{H}$ is defined by $\lambda_{H} u(p, s) = \langle 0 | \eta(0) | H(p, s) \rangle$,
where $u(p, s)$ is the Dirac spinor of the hadron.

Define
\begin{align}
    \label{eq:rho}
    \frac{\Im C_{1}(q)}{\pi}
    & =
    \rho_{1, 1}(q^{2}) \slashed{q}
    +
    \rho_{2, 1}(q^{2})
    \, ,
    \\
    \frac{\Im C_{i}(q)}{\pi}
    & =
    \rho_{1, i}(q^{2}) \slashed{q}
    +
    \rho_{2, i}(q^{2})
    \, ,
\end{align}
and employ the quark-hadron duality and Borel transformation,
we obtain a sum rule corresponding to $\Pi_{1}(q^{2})$ \cite{Ioffe:1981kw}
\begin{align}
    \label{eq:sum-1}
    \lambda_{H}^{2}
    e^{- \frac{m_{H}^{2}}{m_{B}^{2}}}
    & =
    \int_{s_{th}}^{s_{0}} \mathrm{d} s \,
    \rho_{1, 1}(s)
    e^{- \frac{s}{m_{B}^{2}}}
    \nonumber \\
    & \phantom{= {}} +
    \sum_{i}
    \langle O_{i} \rangle
    \int_{s_{th}}^{\infty} \mathrm{d} s \,
    \rho_{1, i}(s)
    e^{- \frac{s}{m_{B}^{2}}}
    \, ,
\end{align}
where $s_{0}$ is the threshold parameter,
and $m_{B}$ is the Borel parameter,
which are introduced in the quark-hadron duality and Borel transformation respectively.
One can also obtain a similar sum rule corresponding to $\Pi_{2}(q^{2})$,
but we will not discuss it in this paper.

To obtain the baryon mass,
we differentiate both sides of Eq.~(\ref{eq:sum-1}) with respect to $- m_{B}^{- 2}$ and solve for $m_{H}^{2}$,
which results in
\begin{align}
    \label{eq:mass}
    m_{H}^{2}
    & =
    \nonumber \\
    &
    \frac{
        \int_{s_{th}}^{s_{0}} \mathrm{d} s \,
        \rho_{1, 1}(s)
        s
        e^{- \frac{s}{m_{B}^{2}}}
        +
        \sum_{i}
        \langle O_{i} \rangle
        \int_{s_{th}}^{\infty} \mathrm{d} s \,
        \rho_{1, i}(s)
        s
        e^{- \frac{s}{m_{B}^{2}}}
    }{
        \int_{s_{th}}^{s_{0}} \mathrm{d} s \,
        \rho_{1, 1}(s)
        e^{- \frac{s}{m_{B}^{2}}}
        +
        \sum_{i}
        \langle O_{i} \rangle
        \int_{s_{th}}^{\infty} \mathrm{d} s \,
        \rho_{1, i}(s)
        e^{- \frac{s}{m_{B}^{2}}}
    }
    \, .
\end{align}
In this paper, as a good approximation, we only keep vacuum condensates up to dimension 4,
\begin{equation}
    \langle O_{i} \rangle
    \in
    \left\{
        \langle \overline{q}_{j}^{a} q_{j}^{a} \rangle,
        \langle g_{s}^{2} G_{\mu \nu}^{a} G^{a \mu \nu} \rangle
    \right\}
    \, ,
\end{equation}
and evaluate $\rho_{1, \langle \overline{q} q \rangle}$ up to $\mathcal{O}(m_{q})$.
Contributions of higher dimensional operators are power suppressed and thus can be neglected
(See App.~(\ref{sec:ope}) for more discussions on higher dimensional operators).

\sect{Baryon Currents}
The most general current of baryon containing two identical heavy quarks is
\begin{equation}
    \label{eq:current}
    \epsilon^{a b c} \left( Q^{a} C \Gamma_{1} Q^{b} \right) \Gamma_{2} q^{c}
    \, ,
\end{equation}
where
$Q$ is the heavy quark with mass $m_{Q}$,
while $q$ is the light quark with mass $m_{q}$.
$\epsilon^{a b c}$ is the antisymmetric matrix in color space,
$C$ is the charge conjugation matrix,
and $\Gamma_{1}$ and $\Gamma_{2}$ are Dirac matrices with possible Lorentz indices suppressed.
Spinor indices are contracted within the bracket,
and therefore transposing the bracket part should keep the current intact.
Note that $C^{T} = - C$, one can see that $\Gamma_{1}$ can only be $\gamma_{\mu}$ or $\sigma_{\mu \nu}$ \cite{Ioffe:1981kw}.
For a $J^{P} = \frac{1}{2}^{+}$ baryon, there are only two possible currents
\begin{align}
    \label{eq:eta1}
    \eta_{1}
    & =
    \epsilon^{a b c} \left( Q^{a} C \gamma_{\mu} Q^{b} \right) \gamma^{\mu} \gamma^{5} q^{c}
    \, ,
    \\
    \label{eq:eta2}
    \eta_{2}
    & =
    \epsilon^{a b c} \left( Q^{a} C \sigma_{\mu \nu} Q^{b} \right) \sigma^{\mu \nu} i \gamma^{5} q^{c}
    \, ,
\end{align}
where $\eta_{1}$ corresponds to the Ioffe current \cite{Ioffe:1981kw} if we take $Q$ as $u$ quark and $q$ as $d$ quark.
It is well known that $\eta_{1}$ and $\eta_{2}$ are renormcovariant \cite{Ioffe:1982ce},
\begin{equation}
    \frac{\mathrm{d}}{\mathrm{d} \ln \mu^{2}}
    \begin{pmatrix}
        \eta_{1}
        \\
        \eta_{2}
    \end{pmatrix}
    =
    \begin{pmatrix}
        \gamma_{1} & 0
        \\
        0 & \gamma_{2}
    \end{pmatrix}
    \begin{pmatrix}
        \eta_{1}
        \\
        \eta_{2}
    \end{pmatrix}
    \, .
\end{equation}
Thus it is advantageous to work with these currents when calculating the NLO correction.
There exist other choices of current \cite{Chung:1981cc, Bagan:1992za, Leinweber:1995fn}, which
can be expressed by $\eta_{1}$ and $\eta_{2}$ with the help of Fierz identity,
\begin{align}
    \label{eq:mix}
    \eta_{\text{mix}}
    & =
    \epsilon^{a b c}
    \left[
        \left( Q^{a} C \gamma^{5} q^{b} \right) Q^{c}
        +
        b
        \left( Q^{a} C q^{b} \right) \gamma^{5} Q^{c}
    \right]
    \nonumber \\
    & =
    \frac{b - 1}{4} \eta_{1} + i \frac{b + 1}{8} \eta_{2}
    \, ,
\end{align}
where $b$ is a complex mixing parameter.

\sect{NLO Correction to $C_{1}$}
It is known that $C_{1}$ and $C_{i}$ can be calculated perturbatively,
and results at LO are available in \cite{Bagan:1992za, Narison:2010py}.
Among them, the most important one is $C_{1}$,
because all other coefficients will be multiplied by higher dimensional operators which are power suppressed.
Thus the main theoretical uncertainty is due to NLO correction to $C_{1}$.

\begin{figure}[ht]
    \centering
    \includegraphics[width=0.9\linewidth]{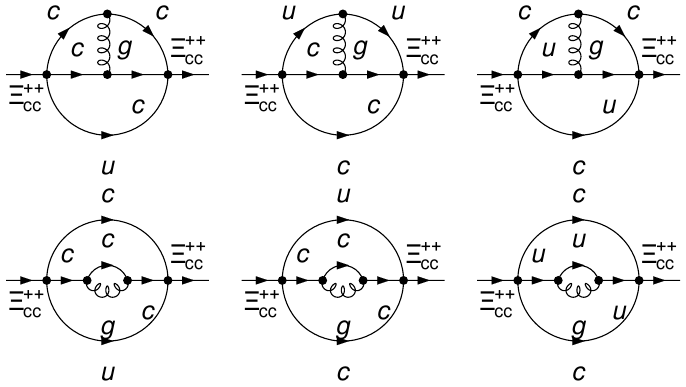}
    \caption{
        NLO Feynman diagrams for $C_{1}$.
        External legs are amputated.
    }
    \label{fig:amp}
\end{figure}

\begin{figure}[ht]
    \centering
    \includegraphics[width=0.9\linewidth]{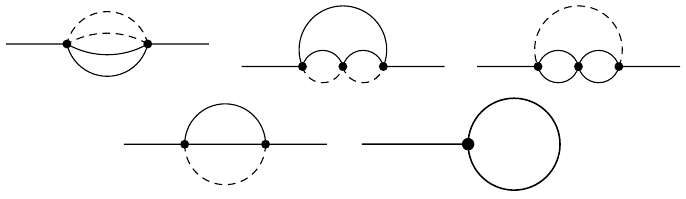}
    \caption{
        Topologies of master integrals,
        where solid and dashed lines denote massive and massless propagators respectively.
        External legs are amputated.
    }
    \label{fig:top}
\end{figure}

In order to perform NLO calculation for $C_{1}$,
we use \verb|FeynArts| \cite{Kublbeck:1990xc, Hahn:2000kx} to generate all Feynman diagrams (see Fig.~(\ref{fig:amp})),
and \verb|FeynCalc| \cite{Mertig:1990an, Shtabovenko:2016sxi} to manipulate resulting amplitude.
After these steps, we are left with some three-loop-like scalar integrals.
These integrals can be further simplified by the integration-by-parts (IBP) method \cite{Chetyrkin:1981qh, Laporta:2001dd}.
\verb|FIRE| \cite{Smirnov:2014hma} and \verb|LiteRed| \cite{Lee:2013mka} are used to reduce the full amplitude to
a linear combination of a complete set of 29 master integrals (see Fig.~(\ref{fig:top})),
\begin{equation}
    \label{eq:amp}
    C_{1}^{\text{NLO}}(\varepsilon, q, m_{Q}) = \sum_{k} c_{k}(\varepsilon, q, m_{Q}) I_{k}(\varepsilon, v)
    \, ,
\end{equation}
where $\varepsilon$ is defined by dimension $D = 4 - 2 \varepsilon$,
$v = \sqrt{1 - \frac{4 m_{Q}^{2}}{q^{2}}}$,
and all coefficients $c_{k}$ are purely imaginary.
Note that here $I_{k}$ is defined to be dimensionless.

Since we are only interested in the imaginary part of the two-point function $\Pi(p^{2})$,
we just need to evaluate the corresponding cut diagrams of $I_{k}$.
But evaluating four-body phase space in the presence of two massive particles is still a formidable task.

To proceed, we employ the differential equation method \cite{Henn:2013pwa, Henn:2014qga}
by first differentiating $I_{k}$ with respect to $v$, then reducing the resulting integrals by using IBP,
and obtaining a system of differential equations,
\begin{equation}
    \frac{\mathrm{d} \boldsymbol{I}(\varepsilon, v)}{\mathrm{d} v}
    =
    \boldsymbol{A}(\varepsilon, v) \boldsymbol{I}(\varepsilon, v)
    \, ,
\end{equation}
where $\boldsymbol{I}$ represents the vector of master integrals $I_{k}$,
and $\boldsymbol{A}$ is a $29 \times 29$ matrix.
To solve this differential equation,
we implement algorithm proposed in \cite{Lee:2014ioa} to transform the equation into the so-called $\varepsilon$-form \cite{Henn:2013pwa},
\begin{equation}
    \label{eq:epsilon}
    \frac{\mathrm{d} \boldsymbol{I}'(\varepsilon, v)}{\mathrm{d} v}
    =
    \varepsilon
    \sum_{i} \frac{\boldsymbol{B}_{i}}{v - v_{i}} \boldsymbol{I}'(\varepsilon, v)
    \, ,
\end{equation}
where
$v_{i} \in \left\{ 0, \pm 1, \pm \sqrt{3} i \right\}$,
$\boldsymbol{B}_{i}$ are constant matrices,
and $\boldsymbol{I}'$ is related to $\boldsymbol{I}$ with an invertible linear transformation.
The virtue of this $\varepsilon$-form is that
the right hand side of Eq.~(\ref{eq:epsilon}) is proportional to $\varepsilon$,
which can be easily solved iteratively in terms of Goncharov polylogarithms \cite{Goncharov:2001iea}.
The boundary values of $\boldsymbol{I}(\varepsilon, v)$ at $v = 1$, i.e. $m_{Q} = 0$,
are nothing but massless four-body phase space integrals,
which are very easy to work out.
By evaluating the boundary value $\boldsymbol{I}(\varepsilon, 1)$,
and solving the equation iteratively, we finally obtain $I_{k}$ and finish our calculation.

We find that the Coulombic singularity, which appears as $v \to 0$, does not present in this order.
Then by combining all terms together, infrared divergences are canceled out,
so we only need to deal with ultraviolet divergences.
After performing wavefunction and mass renormalization of quarks ($m_{Q}$ is renormalized in either $\overline{\text{MS}}$ scheme or on-shell scheme),
the remaining ultraviolet divergences can be removed by operator renormalization of $\eta_{1}$ and $\eta_{2}$.
We renormalize them in $\overline{\text{MS}}$ scheme,
of which anomalous dimensions are
\begin{equation}
    \gamma_{1}
    =
    \gamma_{2}
    =
    \frac{\alpha_{s}}{2 \pi}
    \, ,
\end{equation}
which confirm the results in \cite{Peskin:1979mn, Ovchinnikov:1991mu}.

We then get a finite result at NLO.
Our NLO result confirms the massless result \cite{Jamin:1987gq, Ovchinnikov:1991mu} in the limit of $m_{Q} \to 0$.
Our analytical result is listed in App.~(\ref{sec:result}).

\sect{Phenomenology}
In our analysis, we use
\begin{equation}
    \label{eq:eta}
    \eta
    =
    \eta_{1} + \theta \eta_{2}
    \, ,
\end{equation}
with $\theta$ a complex mixing parameter.
We choose following parameters \cite{Olive:2016xmw, Dominguez:1994ce, Bagan:1992za, Dominguez:2014pga, Aoki:2016frl}:
\begin{align}
    \label{eq:pfirst}
    m_{u}(2 \text{~GeV})
    & =
    2.36 \pm 0.24 \text{~MeV}
    \, ,
    \\
    m_{d}(2 \text{~GeV})
    & =
    5.03 \pm 0.26 \text{~MeV}
    \, ,
    \\
    m_{c}^{\overline{\text{MS}}}(m_{c})
    & =
    1.28 \pm 0.03 \text{~GeV}
    \, ,
    \\
    m_{c}^{\text{on-shell}}
    & =
    1.46 \pm 0.07 \text{~GeV}
    \, ,
    \\
    \langle \overline{q} q \rangle(2 \text{~GeV})
    & =
    - \left( 0.280 \pm 0.017 \text{~GeV} \right)^{3}
    \, ,
    \\
    \langle g_{s}^{2} G G \rangle
    & =
    4 \pi^{2} (0.037 \pm 0.015) \text{~GeV}^{4}
    \, ,
    \label{eq:plast}
\end{align}
and $\alpha_{s}(m_{Z} = 91.1876 \text{~GeV}) = 0.1181$.
The $m_{c}^{\text{on-shell}}$ comes from the QCD sum rules analysis of $J / \psi$ spectrum,
in which the mass renormalization scheme and the truncation order of $\alpha_{s}$ of $C_{1}$ are the same as ours.
Thus it is consistent to use this on-shell quark mass in our analysis.
According to Eq.~(\ref{eq:mass}),
the evolution of the current $\eta$ is irrelevant to the estimation of hadron mass,
thus we do not include it in our analysis.
We use two-loop running for the coupling constant $\alpha_{s}$ and heavy quark mass $m_{Q}$.
The vacuum condensates are evolved according to their one-loop anomalous dimensions:
$\gamma_{\langle \overline{q} q \rangle} = - \gamma_{m_{q}}$
and
$\gamma_{\langle g_{s}^{2} G G \rangle} = 0$ \cite{Albuquerque:2013ija}.
In the following, unless otherwise stated,
we choose central values for all parameters,
set renormalization scale $\mu = m_{B}$ \cite{Shifman:1978bx, Bertlmann:1981he},
and choose ${\overline{\text{MS}}}$ scheme for heavy quark mass renormalization.

In Eq.~(\ref{eq:mass}), the baryon mass $m_{H}$ depends on two parameters: $m_{B}$ and $s_{0}$.
In order to obtain a reliable result,
we should keep $m_{B}$ inside the so-called Borel window to ensure the validity of OPE,
and the choice of $s_{0}$ should ensure the ground-state pole contribution domination.
Since $m_{H}$ is a property of hadron, it does not depend on $m_{B}$ and $s_{0}$,
thus within the valid parameter space (we shall call it ``window'' hereafter),
we should find the region in which $m_{H}$ depends weakly on $m_{B}$ and $s_{0}$.
$m_{H}$ in this region is considered to be the estimated hadron mass in QCD sum rules.

\begin{table*}[ht]
    \caption{
        Parameters of plateau and predictions for $m_{\Xi_{cc}^{++}}$ in different mixing and mass renormalization schemes.
    }
    \begin{ruledtabular}
        \begin{tabular}{ccccccccc}
            $\theta$ & $m_{Q}$ scheme & Order & $m_{B}^{2} \text{~(GeV$^{2}$)}$ & $s_{0} \text{~(GeV$^{2}$)}$ & $m_{\Xi_{cc}^{++}} \text{~(GeV)}$ & Error from $m_{B}^{2}$ & Error from $s_{0}$ & Error from $m_{Q}$
            \\ \hline
            \multirow{2}{*}{$0.018 i$} & \multirow{2}{*}{$\overline{\text{MS}}$} & LO & $2.0 \pm 0.3$ & $17 \pm 2$ & $3.57_{-0.11}^{+0.08}$ & ${-0.00} \; {+0.01}$ & ${-0.09} \; {+0.07}$ & ${-0.05} \; {+0.05}$
            \\
            & & NLO & $1.7 \pm 0.3$ & $17 \pm 2$ & $3.66_{-0.10}^{+0.08}$ & ${-0.01} \; {+0.01}$ & ${-0.08} \; {+0.05}$ & ${-0.05} \; {+0.05}$
            \\ \hline
            \multirow{2}{*}{$0.018 i$} & \multirow{2}{*}{on-shell} & LO & $1.7 \pm 0.3$ & $17 \pm 2$ & $3.83_{-0.14}^{+0.13}$ & ${-0.03} \; {+0.00}$ & ${-0.09} \; {+0.07}$ & ${-0.10} \; {+0.10}$
            \\
            & & NLO & $1.4 \pm 0.3$ & $17 \pm 2$ & $3.65_{-0.14}^{+0.11}$ & ${-0.07} \; {+0.05}$ & ${-0.08} \; {+0.05}$ & ${-0.10} \; {+0.09}$
            \\ \hline
            \multirow{2}{*}{$- \frac{i}{3}$} & \multirow{2}{*}{$\overline{\text{MS}}$} & LO & $4.4 \pm 0.3$ & $23 \pm 2$ & $3.81_{-0.11}^{+0.10}$ & ${-0.04} \; {+0.04}$ & ${-0.10} \; {+0.08}$ & ${-0.03} \; {+0.03}$
            \\
            & & NLO & $4.0 \pm 0.3$ & $23 \pm 2$ & $3.86_{-0.11}^{+0.10}$ & ${-0.05} \; {+0.04}$ & ${-0.09} \; {+0.08}$ & ${-0.03} \; {+0.03}$
        \end{tabular}
    \end{ruledtabular}
    \label{tab:result}
\end{table*}

\begin{figure}[ht]
    \centering
    \begin{subfigure}[ht]{0.6\linewidth}
        \includegraphics[width=\linewidth]{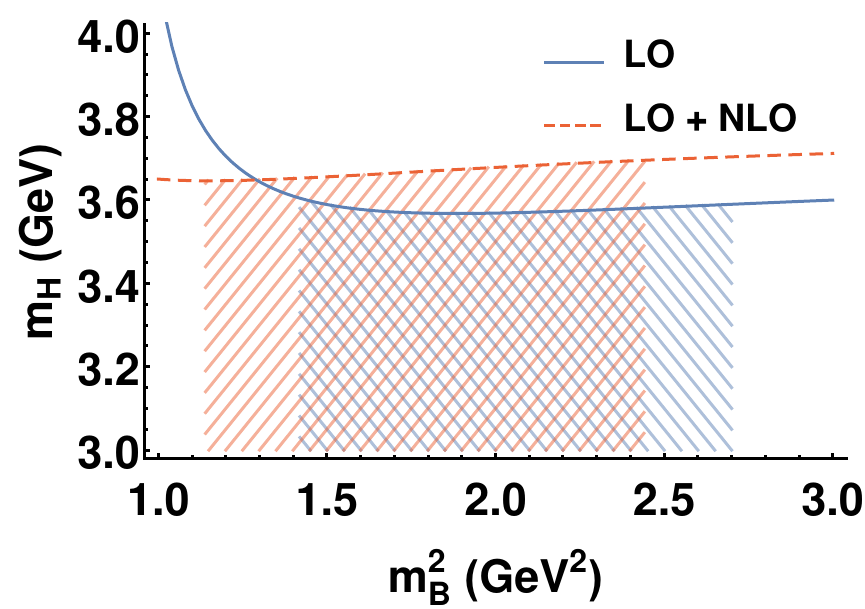}
    \end{subfigure}
    \\
    \begin{subfigure}[ht]{0.6\linewidth}
        \includegraphics[width=\linewidth]{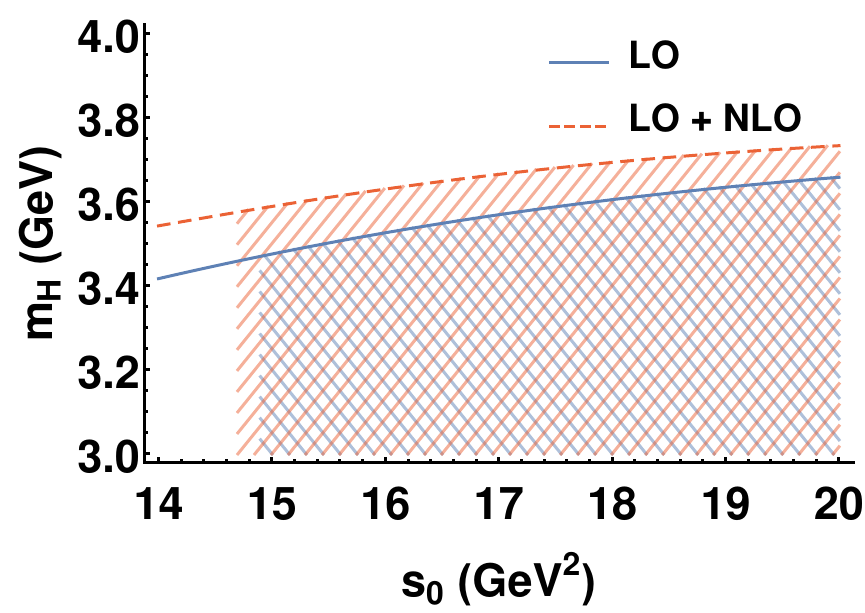}
    \end{subfigure}
    \caption{
        Prediction of $m_{\Xi_{cc}^{++}}$ as a function of $m_{B}^{2}$ and $s_{0}$.
        Shadows correspond to windows defined by Eq.~(\ref{eq:window}).
    }
    \label{fig:msbar}
\end{figure}

We define relative contributions of condensates and continuum spectrum as
\begin{align}
    r_{i}
    & =
    \frac{
        \langle O_{i} \rangle
        \int_{s_{th}}^{\infty} \mathrm{d} s \,
        \rho_{1, i}(s)
        e^{- \frac{s}{m_{B}^{2}}}
    }{
        \int_{s_{th}}^{\infty} \mathrm{d} s \,
        \rho_{1, 1}(s)
        e^{- \frac{s}{m_{B}^{2}}}
    }
    \, ,
    \\
    r_{\text{cont.}}
    & =
    \frac{
        \int_{s_{0}}^{\infty} \mathrm{d} s \,
        \rho_{1, 1}(s)
        e^{- \frac{s}{m_{B}^{2}}}
    }{
        \int_{s_{th}}^{\infty} \mathrm{d} s \,
        \rho_{1, 1}(s)
        e^{- \frac{s}{m_{B}^{2}}}
    }
    \, ,
\end{align}
and impose the following constraints on our sum rule
\begin{equation}
    \label{eq:window}
    \left| r_{i} \right| \le 30 \%
    \, ,
    \quad
    \left| \sum_{i} r_{i} \right| \le 30 \%
    \, ,
    \quad
    \left| r_{\text{cont.}} \right| \le 30 \%
    \, .
\end{equation}
We find that with mixing parameter $\theta = 0.018 i$,
we can obtain a very stable plateau of $m_{B}$ and $s_{0}$, as shown in Fig.~(\ref{fig:msbar}).
Note, however, that QCD sum rules alone cannot tell which mixing current is the most suitable one for QCD sum rules analysis.
For example, there is a family of mixing parameters that can yield similar good plateau of $m_{B}$ and $s_{0}$, and similar estimation of $m_{H}$.
We also provide another set of results by choosing $\theta = - \frac{i}{3}$,
which corresponds to the mixing used in \cite{Bagan:1992za}.

\begin{figure}[ht]
    \centering
    \includegraphics[width=0.6\linewidth]{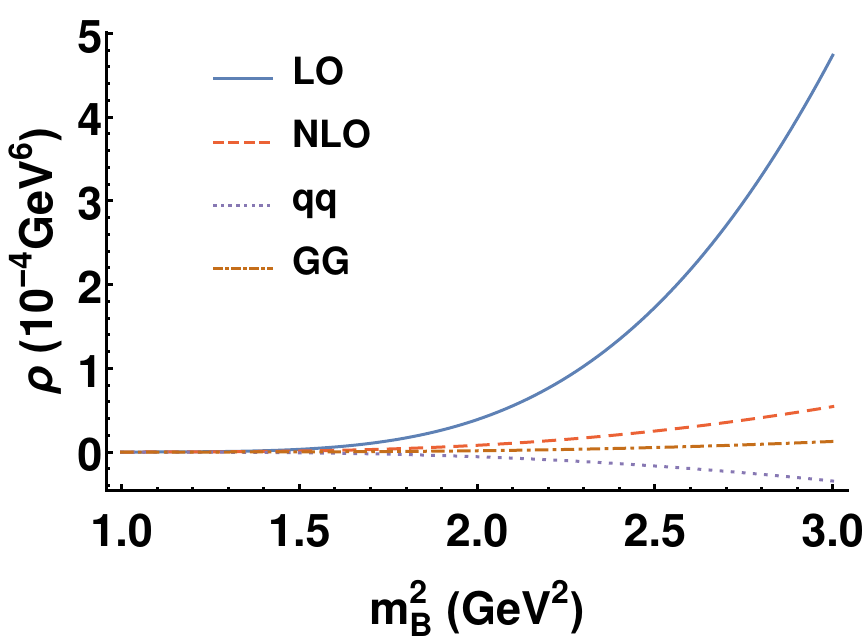}
    \caption{
        Contributions of various terms on the right hand side of Eq.~(\ref{eq:sum-1}).
    }
    \label{fig:ope}
\end{figure}

\begin{figure}[ht]
    \centering
    \includegraphics[width=0.6\linewidth]{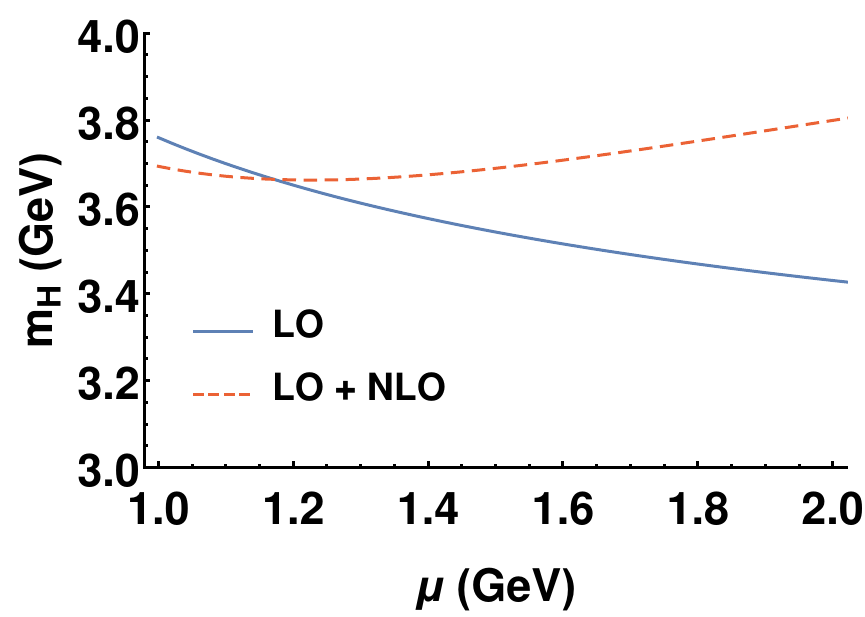}
    \caption{
        Prediction of $m_{\Xi_{cc}^{++}}$ as a function of $\mu$.
    }
    \label{fig:scale}
\end{figure}

The relative importance of each term in OPE is shown in Fig.~(\ref{fig:ope}),
where $m_{B}^{2}$ and $s_{0}$ are set to their central values shown in Tab.~(\ref{tab:result}).
We find that the NLO correction has an important contribution.
In the $m_{Q}^{\overline{\text{MS}}}$ scheme,
the ratio of NLO correction to LO is about $29 \%$ ($19 \%$) for $\theta = 0.018 i$ ($\theta = - \frac{i}{3}$).
While in the $m_{Q}^{\text{on-shell}}$ scheme, this ratio reaches $233 \%$ for $\theta = 0.018 i$,
signaling the bad convergence of perturbative expansion,
which is the reason why we choose $\overline{\text{MS}}$ scheme by default.
Nevertheless, with NLO correction,
the difference of predicted $m_{\Xi_{cc}^{++}}$ between $\overline{\text{MS}}$ scheme and on-shell scheme for $m_{Q}$ is substantially reduced.
As shown in Tab.~(\ref{tab:result}),
the mass differences obtained from LO and $\text{LO} + \text{NLO}$ results are $0.27 \text{~GeV}$ and $0.01 \text{~GeV}$, respectively.
Thus NLO correction largely reduces the scheme dependence.

To study the renormalization scale $\mu$ dependence, we fix all other parameters by their default choices (or central values) and freely vary $\mu$.
The variation of $m_{\Xi_{cc}^{++}}$ with respect to $\mu$ is shown in Fig.~(\ref{fig:scale}).
We find the scale dependence is much weaker when NLO correction is included.
More precisely, the error of $m_{\Xi_{cc}^{++}}$ induced by $\mu = m_{B} \pm 0.2 \text{~GeV}$ is $_{-0.08}^{+0.06} \text{~GeV}$ and $_{-0.01}^{+0.03} \text{~GeV}$ in LO and $\text{LO} + \text{NLO}$, respectively.

Our final results for $m_{\Xi_{cc}^{++}}$ are shown in Tab.~(\ref{tab:result}).
Errors of $m_{B}^{2}$, $s_{0}$ and parameters listed in Eq.~(\ref{eq:pfirst})-(\ref{eq:plast}) are used to determine the error of $m_{\Xi_{cc}^{++}}$.
We find that our NLO result is consistent with the LHCb measurement.
As a comparison, we also list the results with $m_{Q}^{\text{on-shell}}$ renormalization scheme or with $\theta = - \frac{i}{3}$.
We find that all plots above are almost unchanged when changing $m_{q}$ from $m_{u}$ to $m_{d}$,
thus our prediction of the mass of $\Xi_{cc}^{+} (ccd)$ is almost the same as that of $\Xi_{cc}^{++}(ccu)$.

\sect{Summary}
The NLO calculation for hadrons with massive quarks in QCD sum rules is important but hard to carry out.
With the help of recent development of multi-loop calculation technique,
we are able to analytically calculate the NLO perturbative correction to the imaginary part of the two-point correlation function of
$J^{P} = \frac{1}{2}^{+}$ baryon current with two identical heavy quarks.
We apply our result to the QCD sum rules analysis of newly discovered baryon $\Xi_{cc}^{++}$ by LHCb \cite{Aaij:2017ueg}.
The QCD sum rules estimation of $m_{\Xi_{cc}^{++}}$ is $3.66_{-0.10}^{+0.08} \text{~GeV}$, which is consistent with the LHCb measurement within uncertainties.
By comparing LO with $\text{LO} + \text{NLO}$ results, we find the NLO perturbative correction substantially reduces $m_{Q}$ renormalization scheme dependence and renormalization scale $\mu$ dependence,
thus makes the theoretical uncertainties under better control.

\begin{acknowledgments}
We thank H. X. Chen and S. L. Zhu for many useful communications and discussions.
The work is supported in part by
the National Natural Science Foundation of China (Grants No. 11475005 and No. 11075002),
and the National Key Basic Research Program of China (No. 2015CB856700).
\end{acknowledgments}

\appendix

\sect{Analytical Result}
\label{sec:result}
We calculate various spectrum densities of the current defined in Eq.~(\ref{eq:eta}).
The corresponding LO spectrum densities, defined in Eq.~(\ref{eq:rho}), are
\begin{widetext}
\begin{align}
    \rho_{1, 1}^{\text{LO}}
    & =
    \frac{1}{2048 \pi^{4}} q^{4}
    \left[
        2 v (9 v^{6} - 9 v^{4} + 31 v^{2} - 15)
        +
        3 (v^{2} - 1)^{3} (3 v^{2} + 5)
        \ln \left( \frac{1 - v}{1 + v} \right)
    \right]
    \nonumber \\
    & \phantom{= {}}
    +
    \frac{3}{1024 \pi^{4}} q^{4}
    |\theta|^{2}
    \left[
        2 v (3 v^{6} - 11 v^{4} + 69 v^{2} - 45)
        +
        3 (v^{2} - 1)^{2} (v^{4} - 2 v^{2} - 15)
        \ln \left( \frac{1 - v}{1 + v} \right)
    \right]
    \, ,
    \\
    \rho_{2, 1}^{\text{LO}}
    & =
    \frac{3}{128 \pi^{4}} q^{4} m_{Q}
    \Im \theta
    \left[
        2 v (v^{2} + 3) (3 v^{2} - 5)
        +
        3 (v^{2} - 1) (v^{4} + 2 v^{2} + 5)
        \ln \left( \frac{1 - v}{1 + v} \right)
    \right]
    \, ,
    \\
    \rho_{1, \langle \overline{q} q \rangle}^{\text{LO}}
    & =
    \frac{3}{\pi^{2}} m_{Q} \Im \theta
    v
    -
    \frac{1}{32 \pi^{2}} m_{q}
    \frac{v^{4} - 6 v^{2} - 3}{v}
    -
    \frac{3}{8 \pi^{2}} m_{q} |\theta|^{2}
    \frac{v^{4} - 4 v^{2} - 1}{v}
    \, ,
    \\
    \rho_{2, \langle \overline{q} q \rangle}^{\text{LO}}
    & =
    \frac{1}{8 \pi^{2}} q^{2}
    v (v^{2} - 3)
    +
    \frac{3}{2 \pi^{2}} q^{2} |\theta|^{2}
    v (v^{2} - 1)
    -
    \frac{3}{4 \pi^{2}} m_{Q} m_{q} \Im \theta
    \frac{3 v^{2} + 1}{v}
    \, ,
    \\
    \rho_{1, \langle g_{s}^{2} G G \rangle}^{\text{LO}}
    & =
    \frac{1}{512 \pi^{4}}
    \left[
        2 v (v^{2} + 1)
        +
        (v^{2} - 1)^{2}
        \ln \left( \frac{1 - v}{1 + v} \right)
    \right]
    -
    \frac{1}{256 \pi^{4}}
    |\theta|^{2}
    \left[
        2 v (v^{2} + 1)
        +
        (v^{2} - 1)^{2}
        \ln \left( \frac{1 - v}{1 + v} \right)
    \right]
    \, ,
    \\
    \rho_{2, \langle g_{s}^{2} G G \rangle}^{\text{LO}}
    & =
    -
    \frac{1}{64 \pi^{4}} m_{Q}
    \Im \theta
    \left[
        \frac{2 v (3 v^{2} - 11)}{v^{2} - 1}
        +
        (v^{2} + 11)
        \ln \left( \frac{1 - v}{1 + v} \right)
    \right]
    \, .
\end{align}
\end{widetext}
With the help of Eq.~(\ref{eq:mix}),
our result confirms previous calculations \cite{Bagan:1992za, Narison:2010py}.

The NLO spectrum densities of perturbative contribution in $\overline{\text{MS}}$ scheme,
with $m_{Q}$ also renormalized in $\overline{\text{MS}}$ scheme, are
\begin{align}
    \left. \rho_{1, 1}^{\text{NLO}} \right|_{m_{Q}^{\overline{\text{MS}}}}
    & =
    \frac{\alpha_{s}}{2 \pi}
    \left[
        2
        \rho_{1, 1}^{\text{LO}}
        \ln \left( \frac{\mu^{2}}{q^{2}} \right)
        +
        2
        \rho_{a}^{m_{Q}}
        \ln \left( \frac{\mu^{2}}{q^{2}} \right)
        +
        \rho_{a}
    \right]
    \, ,
    \\
    \left. \rho_{2, 1}^{\text{NLO}} \right|_{m_{Q}^{\overline{\text{MS}}}}
    & =
    \frac{\alpha_{s}}{2 \pi}
    \left[
        2
        \rho_{2, 1}^{\text{LO}}
        \ln \left( \frac{\mu^{2}}{q^{2}} \right)
        +
        2
        \rho_{b}^{m_{Q}}
        \ln \left( \frac{\mu^{2}}{q^{2}} \right)
        +
        \rho_{b}
    \right]
    \, ,
\end{align}
where $\rho_{a}^{m_{Q}}$ and $\rho_{b}^{m_{Q}}$ come from $m_{Q}$ renormalization
\begin{equation}
    \rho_{a}^{m_{Q}}
    =
    m_{Q} \frac{\partial}{\partial m_{Q}} \rho_{1, 1}^{\text{LO}}
    \, ,
    \quad
    \rho_{b}^{m_{Q}}
    =
    m_{Q} \frac{\partial}{\partial m_{Q}} \rho_{2, 1}^{\text{LO}}
    \, .
\end{equation}
The analytical expressions of $\rho_{a}$ and $\rho_{b}$ will be presented later.
The differences between $m_{Q}^{\text{on-shell}}$ scheme and $m_{Q}^{\overline{\text{MS}}}$ scheme are
\begin{align}
    \label{eq:diff-1}
    \left. \rho_{1, 1}^{\text{NLO}} \right|_{m_{Q}^{\text{on-shell}}}
    & =
    \left. \rho_{1, 1}^{\text{NLO}} \right|_{m_{Q}^{\overline{\text{MS}}}}
    -
    \frac{\alpha_{s}}{2 \pi}
    \left[ \frac{8}{3} + 2 \ln \left( \frac{\mu^{2}}{m_{Q}^{2}} \right) \right]
    \rho_{a}^{m_{Q}}
    \, ,
    \\
    \label{eq:diff-2}
    \left. \rho_{2, 1}^{\text{NLO}} \right|_{m_{Q}^{\text{on-shell}}}
    & =
    \left. \rho_{2, 1}^{\text{NLO}} \right|_{m_{Q}^{\overline{\text{MS}}}}
    -
    \frac{\alpha_{s}}{2 \pi}
    \left[ \frac{8}{3} + 2 \ln \left( \frac{\mu^{2}}{m_{Q}^{2}} \right) \right]
    \rho_{b}^{m_{Q}}
    \, .
\end{align}
Note that in the $m_{Q}^{\text{on-shell}}$ scheme,
the logarithms coming from $m_{Q}$ renormalization are completely canceled out,
only the logarithms proportional to $\rho^{\text{LO}}$ remain,
which come from the quark wavefunction renormalization and baryon operator renormalization.
Eq.~(\ref{eq:diff-1}) and Eq.~(\ref{eq:diff-2}) are just the consequences of changing renormalization scheme.
To show this explicitly,
we first replace all $m_{Q}^{\text{on-shell}}$ by $m_{Q}^{\overline{\text{MS}}}$ in $\rho^{\text{LO}}$ and $\rho^{\text{NLO}}$ in the $m_{Q}^{\text{on-shell}}$ scheme
\begin{align}
    &
    m_{Q}^{\text{on-shell}}
    \nonumber \\
    = {} &
    m_{Q}^{\overline{\text{MS}}}
    \left( 1 + \frac{\alpha_{s}}{2 \pi} \left[ \frac{8}{3} + 2 \ln \left( \frac{\mu^{2}}{\left( m_{Q}^{\overline{\text{MS}}} \right)^{2}} \right) \right] + \mathcal{O}(\alpha_{s}^{2}) \right)
    \, .
\end{align}
Then we expand $\rho^{\text{LO}}$ and $\rho^{\text{NLO}}$ up to $\mathcal{O}(\alpha_{s})$.
We take $\rho_{1, 1}$ for example.
For $\rho_{1, 1}^{\text{LO}}$ we have
\begin{align}
    &
    \rho_{1, 1}^{\text{LO}}(m_{Q}^{\text{on-shell}})
    \nonumber \\
    = {} &
    \rho_{1, 1}^{\text{LO}}(m_{Q}^{\overline{\text{MS}}})
    +
    \frac{\alpha_{s}}{2 \pi}
    \left[ \frac{8}{3} + 2 \ln \left( \frac{\mu^{2}}{\left( m_{Q}^{\overline{\text{MS}}} \right)^{2}} \right) \right]
    \rho_{a}^{m_{Q}}(m_{Q}^{\overline{\text{MS}}})
    \nonumber \\
    \phantom{= {}} & +
    \mathcal{O}(\alpha_{s}^{2})
    \, ,
\end{align}
and for $\rho_{1, 1}^{\text{NLO}}$
\begin{align}
    \left. \rho_{1, 1}^{\text{NLO}} \right|_{m_{Q}^{\text{on-shell}}}(m_{Q}^{\text{on-shell}})
    & =
    \left. \rho_{1, 1}^{\text{NLO}} \right|_{m_{Q}^{\text{on-shell}}}(m_{Q}^{\overline{\text{MS}}})
    \nonumber \\
    & \phantom{= {}} +
    \mathcal{O}(\alpha_{s}^{2})
    \, .
\end{align}
Combining them together, we obtain
\begin{align}
    &
    \rho_{1, 1}^{\text{LO}}(m_{Q}^{\text{on-shell}})
    +
    \left. \rho_{1, 1}^{\text{NLO}} \right|_{m_{Q}^{\text{on-shell}}}(m_{Q}^{\text{on-shell}})
    \nonumber \\
    = {} &
    \rho_{1, 1}^{\text{LO}}(m_{Q}^{\overline{\text{MS}}})
    +
    \left. \rho_{1, 1}^{\text{NLO}} \right|_{m_{Q}^{\text{on-shell}}}(m_{Q}^{\overline{\text{MS}}})
    \nonumber \\
    \phantom{= {}} & +
    \frac{\alpha_{s}}{2 \pi}
    \left[ \frac{8}{3} + 2 \ln \left( \frac{\mu^{2}}{\left( m_{Q}^{\overline{\text{MS}}} \right)^{2}} \right) \right]
    \rho_{a}^{m_{Q}}(m_{Q}^{\overline{\text{MS}}})
    \nonumber \\
    \phantom{= {}} & +
    \mathcal{O}(\alpha_{s}^{2}(\mu))
    \, .
\end{align}
Since the renormalized amplitude should not depend on renormalization scheme,
we thus obtain Eq.~(\ref{eq:diff-1}).
For $\rho_{2, 1}$, the result is similar,
all we need to do is substituting
$\rho_{1, 1}^{\text{LO}}$, $\rho_{1, 1}^{\text{NLO}}$, and $\rho_{a}^{m_{Q}}$
in above expressions with
$\rho_{2, 1}^{\text{LO}}$, $\rho_{2, 1}^{\text{NLO}}$, and $\rho_{b}^{m_{Q}}$,
respectively.

As a check,
we can verify that in the $m_{Q}^{\overline{\text{MS}}}$ scheme,
the $\mu$ dependence of $m_{Q}$ in $\rho^{\text{LO}}$ is canceled by corresponding logarithms in $\rho^{\text{NLO}}$ to $\mathcal{O}(\alpha_{s})$.
To show this explicitly,
we first replace all $m_{Q}(\mu)$ by $m_{Q}(\lambda)$ in $\rho^{\text{LO}}$ and $\rho^{\text{NLO}}$ in the $m_{Q}^{\overline{\text{MS}}}$ scheme
\begin{equation}
    m_{Q}(\mu)
    =
    m_{Q}(\lambda)
    \left( 1 - \frac{\alpha_{s}(\mu)}{\pi} \ln \left( \frac{\mu^{2}}{\lambda^{2}} \right) + \mathcal{O}(\alpha_{s}^{2}(\mu)) \right)
    \, ,
\end{equation}
where $\lambda$ is another scale that differs from $\mu$.
Then we expand $\rho^{\text{LO}}$ and $\rho^{\text{NLO}}$ up to $\mathcal{O}(\alpha_{s}(\mu))$,
and the $\mu$ dependence of $m_{Q}$ should cancel out up to $\mathcal{O}(\alpha_{s}(\mu))$.
We take $\rho_{1, 1}$ for example.
For $\rho_{1, 1}^{\text{LO}}$ we have
\begin{align}
    \rho_{1, 1}^{\text{LO}}(m_{Q}(\mu))
    & =
    \rho_{1, 1}^{\text{LO}}(m_{Q}(\lambda))
    \nonumber \\
    & \phantom{= {}} -
    \frac{\alpha_{s}(\mu)}{\pi}
    \ln \left( \frac{\mu^{2}}{\lambda^{2}} \right)
    \rho_{a}^{m_{Q}}(m_{Q}(\lambda))
    \nonumber \\
    & \phantom{= {}} +
    \mathcal{O}(\alpha_{s}^{2}(\mu))
    \, ,
\end{align}
and for $\rho_{1, 1}^{\text{NLO}}$
\begin{align}
    \rho_{1, 1}^{\text{NLO}}(\mu, \alpha_{s}(\mu), m_{Q}(\mu))
    & =
    \rho_{1, 1}^{\text{NLO}}(\mu, \alpha_{s}(\mu), m_{Q}(\lambda))
    \nonumber \\
    & \phantom{= {}} +
    \mathcal{O}(\alpha_{s}^{2}(\mu))
    \, .
\end{align}
Combining them together, we obtain
\begin{align}
    &
    \rho_{1, 1}^{\text{LO}}(m_{Q}(\mu))
    +
    \rho_{1, 1}^{\text{NLO}}(\mu, \alpha_{s}(\mu), m_{Q}(\mu))
    \nonumber \\
    = {} &
    \rho_{1, 1}^{\text{LO}}(m_{Q}(\lambda))
    \nonumber \\
    \phantom{= {}} & +
    \frac{\alpha_{s}(\mu)}{2 \pi}
    \left[
        2
        \rho_{1, 1}^{\text{LO}}(m_{Q}(\lambda))
        \ln \left( \frac{\mu^{2}}{q^{2}} \right)
        +
        \rho_{a}(m_{Q}(\lambda))
    \right]
    \nonumber \\
    \phantom{= {}} & +
    \frac{\alpha_{s}(\mu)}{2 \pi}
    \left[
        2
        \rho_{a}^{m_{Q}}(m_{Q}(\lambda))
        \ln \left( \frac{\lambda^{2}}{q^{2}} \right)
    \right]
    +
    \mathcal{O}(\alpha_{s}^{2}(\mu))
    \, .
\end{align}
For $\rho_{2, 1}$, the result is similar,
all we need to do is substituting
$\rho_{1, 1}^{\text{LO}}$, $\rho_{1, 1}^{\text{NLO}}$, $\rho_{a}$, and $\rho_{a}^{m_{Q}}$
in above expressions with
$\rho_{2, 1}^{\text{LO}}$, $\rho_{2, 1}^{\text{NLO}}$, $\rho_{b}$, and $\rho_{b}^{m_{Q}}$,
respectively.
Thus we have shown that the $\mu$ dependence of $m_{Q}$ is indeed canceled out.

Now we list $\rho_{a}$ and $\rho_{b}$
\begin{align}
    \rho_{a}
    & =
    q^{4} \left( \sum_{i = 1}^{11} g_{1, i} G_{i} + |\theta|^{2} \sum_{i = 1}^{11} g_{2, i} G_{i} \right)
    \, ,
    \\
    \rho_{b}
    & =
    q^{4} m_{Q} \Im \theta \sum_{i = 1}^{11} g_{2, i} G_{i}
    \, ,
\end{align}
where $G_{i}$ are defined as
\begin{widetext}
\begin{align}
    G_{1}
    & =
    18 G_{0,0,1}(1-v)-27 G_{0,0,2}(1-v)+3 G_{0,2,0}(1-v)-18 G_{0,2,1}(1-v)+24
    G_{0,2,2}(1-v)-24 G_{2,0,0}(1-v)
    \nonumber \\
    & \phantom{= {}}
    +18 G_{2,0,1}(1-v)-3 G_{2,0,2}(1-v)+27
    G_{2,2,0}(1-v)-18 G_{2,2,1}(1-v)+12 G_{1-i \sqrt{3},0,0}(1-v)
    \nonumber \\
    & \phantom{= {}}
    -18 G_{1-i
    \sqrt{3},0,1}(1-v)+15 G_{1-i \sqrt{3},0,2}(1-v)-15 G_{1-i
    \sqrt{3},2,0}(1-v)+18 G_{1-i \sqrt{3},2,1}(1-v)
    \nonumber \\
    & \phantom{= {}}
    -12 G_{1-i
    \sqrt{3},2,2}(1-v)+12 G_{1+i \sqrt{3},0,0}(1-v)-18 G_{1+i
    \sqrt{3},0,1}(1-v)+15 G_{1+i \sqrt{3},0,2}(1-v)
    \nonumber \\
    & \phantom{= {}}
    -15 G_{1+i
    \sqrt{3},2,0}(1-v)+18 G_{1+i \sqrt{3},2,1}(1-v)-12 G_{1+i
    \sqrt{3},2,2}(1-v)-3 \ln (2) G_{0,2}(1-v)+24 \ln (2) G_{2,0}(1-v)
    \nonumber \\
    & \phantom{= {}}
    -27 \ln
    (2) G_{2,2}(1-v)-12 \ln (2) G_{1-i \sqrt{3},0}(1-v)+15 \ln (2) G_{1-i
    \sqrt{3},2}(1-v)-12 \ln (2) G_{1+i \sqrt{3},0}(1-v)
    \nonumber \\
    & \phantom{= {}}
    +15 \ln (2) G_{1+i
    \sqrt{3},2}(1-v)+4 \left(\pi ^2-3 \ln (2)^2\right) G_2(1-v)-2 \left(\pi
    ^2-3 \ln (2)^2\right) G_{1-i \sqrt{3}}(1-v)
    \nonumber \\
    & \phantom{= {}}
    -2 \left(\pi ^2-3 \ln
    (2)^2\right) G_{1+i \sqrt{3}}(1-v)+9 \zeta (3)
    \, ,
    \\
    G_{2}
    & =
    4 G_{0,0,1}(1-v)-6 G_{0,0,2}(1-v)-4 G_{0,1,0}(1-v)+4 G_{0,1,2}(1-v)+6
    G_{0,2,0}(1-v)-4 G_{0,2,1}(1-v)
    \nonumber \\
    & \phantom{= {}}
    +4 G_{2,0,1}(1-v)-6 G_{2,0,2}(1-v)-4
    G_{2,1,0}(1-v)+4 G_{2,1,2}(1-v)+6 G_{2,2,0}(1-v)-4 G_{2,2,1}(1-v)
    \nonumber \\
    & \phantom{= {}}
    +4 \ln (2)
    G_{0,1}(1-v)-6 \ln (2) G_{0,2}(1-v)+4 \ln (2) G_{2,1}(1-v)-6 \ln (2)
    G_{2,2}(1-v)+3 \zeta (3)
    \, ,
    \\
    G_{3}
    & =
    2 G_{0,0}(1-v)-2 G_{2,2}(1-v)-2 \ln (2) G_0(1-v)+\ln (2)^2
    \, ,
    \\
    G_{4}
    & =
    -6 G_{0,0}(1-v)+6 G_{0,2}(1-v)+6 G_{2,0}(1-v)-6 G_{2,2}(1-v)+6 \ln (2)
    G_0(1-v)-6 \ln (2) G_2(1-v)+\pi ^2
    \nonumber \\
    & \phantom{= {}}
    -3 \ln (2)^2
    \, ,
    \\
    G_{5}
    & =
    6 G_{0,2}(1-v)-6 G_{2,0}(1-v)+6 \ln (2) G_2(1-v)+\pi ^2
    \, ,
    \\
    G_{6}
    & =
    G_{1,0}(1-v)-G_{1,2}(1-v)-\ln (2) G_1(1-v)
    \, ,
    \\
    G_{7}
    & =
    4 G_{0,1}(1-v)-4 G_{2,1}(1-v)+\pi ^2
    \, ,
    \\
    G_{8}
    & =
    G_0(1-v)+G_2(1-v)-\ln (2)
    \, ,
    \\
    G_{9}
    & =
    G_0(1-v)-G_2(1-v)-\ln (2)
    \, ,
    \\
    G_{10}
    & =
    G_1(1-v)
    \, ,
    \\
    G_{11}
    & =
    1
    \, ,
\end{align}
$g_{1, i}$ are
\begin{align}
    g_{1, 1}
    & =
    -\frac{(v-1)^3 (v+1)^3 \left(3 v^2+5\right)}{1536 \pi ^4}
    \, ,
    \\
    g_{1, 2}
    & =
    0
    \, ,
    \\
    g_{1, 3}
    & =
    -\frac{101 v^{12}+378 v^{10}-1149 v^8-5300 v^6+1883 v^4-102 v^2+93}{6144
    \pi ^4 \left(v^2+3\right)^2}
    \, ,
    \\
    g_{1, 4}
    & =
    \frac{v \left(9 v^6-9 v^4+31 v^2-15\right)}{1152 \pi ^4}
    \, ,
    \\
    g_{1, 5}
    & =
    \frac{257 v^{12}+1290 v^{10}+1119 v^8-500 v^6-1489 v^4+8394 v^2-879}{18432
    \pi ^4 \left(v^2+3\right)^2}
    \, ,
    \\
    g_{1, 6}
    & =
    -\frac{v^6 \left(v^2-3\right)}{24 \pi ^4}
    \, ,
    \\
    g_{1, 7}
    & =
    -\frac{107 v^{12}+482 v^{10}+153 v^8-532 v^6+269 v^4+2514 v^2-945}{3072
    \pi ^4 \left(v^2+3\right)^2}
    \, ,
    \\
    g_{1, 8}
    & =
    \frac{v \left(39 v^8+16 v^6+130 v^4+888 v^2-81\right)}{768 \pi ^4
    \left(v^2+3\right)}
    \, ,
    \\
    g_{1, 9}
    & =
    \frac{4793 v^{10}+2855 v^8-39174 v^6-8018 v^4+6397 v^2-14469}{36864 \pi ^4
    \left(v^2+3\right)}
    \, ,
    \\
    g_{1, 10}
    & =
    -\frac{v \left(321 v^8+374 v^6-448 v^4+2874 v^2-945\right)}{1152 \pi ^4
    \left(v^2+3\right)}
    \, ,
    \\
    g_{1, 11}
    & =
    \frac{v \left(7361 v^6-12289 v^4+18199 v^2-9863\right)}{18432 \pi ^4}
    \, ,
\end{align}
$g_{2, i}$ are
\begin{align}
    g_{2, 1}
    & =
    -\frac{(v-1)^2 (v+1)^2 \left(v^2-5\right) \left(v^2+3\right)}{256 \pi ^4}
    \, ,
    \\
    g_{2, 2}
    & =
    \frac{(v-1)^2 (v+1)^2}{8 \pi ^4}
    \, ,
    \\
    g_{2, 3}
    & =
    -\frac{11 v^{10}+133 v^8-4482 v^6-2054 v^4+3911 v^2-591}{3072 \pi ^4
    \left(v^2+3\right)}
    \, ,
    \\
    g_{2, 4}
    & =
    \frac{v \left(3 v^6-11 v^4+69 v^2-45\right)}{192 \pi ^4}
    \, ,
    \\
    g_{2, 5}
    & =
    \frac{347 v^{10}-419 v^8-2610 v^6+154 v^4+21287 v^2-12615}{9216 \pi ^4
    \left(v^2+3\right)}
    \, ,
    \\
    g_{2, 6}
    & =
    -\frac{v^2 \left(v^2+2\right) \left(v^4-6 v^2+3\right)}{12 \pi ^4}
    \, ,
    \\
    g_{2, 7}
    & =
    -\frac{137 v^{10}-173 v^8-894 v^6+782 v^4+5605 v^2-3921}{1536 \pi ^4
    \left(v^2+3\right)}
    \, ,
    \\
    g_{2, 8}
    & =
    \frac{v \left(42 v^6-235 v^4+1090 v^2-501\right)}{192 \pi ^4}
    \, ,
    \\
    g_{2, 9}
    & =
    \frac{4823 v^8-19532 v^6-40278 v^4+26932 v^2-9961}{18432 \pi ^4}
    \, ,
    \\
    g_{2, 10}
    & =
    -\frac{v \left(411 v^6-1831 v^4+7069 v^2-3921\right)}{576 \pi ^4}
    \, ,
    \\
    g_{2, 11}
    & =
    \frac{v \left(8111 v^6-29663 v^4+104473 v^2-72697\right)}{9216 \pi ^4}
    \, ,
\end{align}
and finally $g_{3, i}$ are
\begin{align}
    g_{3, 1}
    & =
    -\frac{(v-1) (v+1) \left(v^4+2 v^2+5\right)}{32 \pi ^4}
    \, ,
    \\
    g_{3, 2}
    & =
    -\frac{3 (v-1) (v+1)}{4 \pi ^4}
    \, ,
    \\
    g_{3, 3}
    & =
    -\frac{8 v^{10}+35 v^8+20 v^6+114 v^4+556 v^2-221}{32 \pi ^4
    \left(v^2+3\right)^2}
    \, ,
    \\
    g_{3, 4}
    & =
    \frac{v \left(v^2+3\right) \left(3 v^2-5\right)}{24 \pi ^4}
    \, ,
    \\
    g_{3, 5}
    & =
    \frac{20 v^{10}+155 v^8+500 v^6+666 v^4-704 v^2-1661}{96 \pi ^4
    \left(v^2+3\right)^2}
    \, ,
    \\
    g_{3, 6}
    & =
    -\frac{v^2 \left(v^4-3 v^2+6\right)}{2 \pi ^4}
    \, ,
    \\
    g_{3, 7}
    & =
    -\frac{2 v^{10}+14 v^8+35 v^6+29 v^4-57 v^2-87}{4 \pi ^4
    \left(v^2+3\right)^2}
    \, ,
    \\
    g_{3, 8}
    & =
    \frac{v \left(3 v^6+13 v^4-24 v^2-120\right)}{4 \pi ^4 \left(v^2+3\right)}
    \, ,
    \\
    g_{3, 9}
    & =
    \frac{89 v^8+177 v^6-39 v^4+859 v^2+450}{48 \pi ^4 \left(v^2+3\right)}
    \, ,
    \\
    g_{3, 10}
    & =
    -\frac{v \left(12 v^6+43 v^4-33 v^2-174\right)}{3 \pi ^4
    \left(v^2+3\right)}
    \, ,
    \\
    g_{3, 11}
    & =
    \frac{v \left(137 v^4-9 v^2-314\right)}{24 \pi ^4}
    \, .
\end{align}
\end{widetext}
In our result,
the Goncharov polylogarithm is defined as
\begin{align}
    G_{a_{1}}(z)
    & =
    \int_{0}^{z} \frac{\mathrm{d} t}{t - a_{1}}
    \, ,
    \\
    G_{a_{1}, \ldots, a_{n}}(z)
    & =
    \int_{0}^{z} \frac{\mathrm{d} t}{t - a_{1}} G_{a_{2}, \ldots, a_{n}}(t)
    \, ,
\end{align}
and $G_{0, \ldots, 0}(z) = \frac{\ln^{n} (z)}{n !}$ if for all $a_{i} = 0$.

As another check,
we can verify that our result reduces to the massless result in the limit of $m_{Q} \to 0$.
This limit is easy to obtain since the coefficients of Goncharov polylogarithms have no singularities at $v = 1$,
and Goncharov polylogarithms with non-zero $a_{i}$ themselves vanish trivially when $z \to 0$.
In the massless limit, $\rho_{a}$ and $\rho_{b}$ are
\begin{widetext}
\begin{align}
    \left. \rho_{1, 1}^{\text{NLO}} \right|_{m_{Q}^{\overline{\text{MS}}}, m_{Q} \to 0}
    =
    \left. \rho_{1, 1}^{\text{NLO}} \right|_{m_{Q}^{\text{on-shell}}, m_{Q} \to 0}
    & =
    \frac{\alpha_{s}}{2 \pi}
    q^{4}
    \frac{71 + 12 \ln \left( \frac{\mu^{2}}{q^{2}} \right)}{384 \pi^{4}}
    \left( 1 + 6 |\theta|^{2} \right)
    \, ,
    \\
    \left. \rho_{2, 1}^{\text{NLO}} \right|_{m_{Q}^{\overline{\text{MS}}}, m_{Q} \to 0}
    =
    \left. \rho_{2, 1}^{\text{NLO}} \right|_{m_{Q}^{\text{on-shell}}, m_{Q} \to 0}
    & =
    0
    \, .
\end{align}
\end{widetext}
These results confirm the massless results obtained previously \cite{Jamin:1987gq, Ovchinnikov:1991mu}.
Another interesting limit is the threshold limit $q^{2} \to 4 m_{Q}^{2}$.
After straightforward integration and expansion in $v$,
the leading power terms of $v$ are
\begin{widetext}
\begin{align}
    &
    \left. \rho_{1, 1}^{\text{NLO}} \right|_{m_{Q}^{\text{on-shell}}, q^{2} \to 4 m_{Q}^{2}}
    \nonumber \\
    = {} &
    \frac{\alpha_{s}}{2 \pi}
    (4 m_{Q}^{2})^{2}
    \left[
        \frac{1}{32 \pi^{2}}
        \left( 1 + 4 |\theta|^{2} \right)
        v^{6}
        +
        \frac{2 (1161 + 70 \pi^{2} - 945 \ln 2 - 630 \ln v) + 315 \ln \left( \frac{\mu^{2}}{m_{Q}^{2}} \right)}{7350 \pi^{4}}
        \left( 1 + 4 |\theta|^{2} \right)
        v^{7}
        +
        \mathcal{O}(v^{8})
    \right]
    \, ,
    \\
    &
    \left. \rho_{2, 1}^{\text{NLO}} \right|_{m_{Q}^{\text{on-shell}}, q^{2} \to 4 m_{Q}^{2}}
    \nonumber \\
    = {} &
    \frac{\alpha_{s}}{2 \pi}
    (4 m_{Q}^{2})^{2}
    m_{Q}
    \Im \theta
    \left[
        -
        \frac{1}{4 \pi^{2}}
        v^{6}
        -
        \frac{4 \left( 2 (1161 + 70 \pi^{2} - 945 \ln 2 - 630 \ln v) + 315 \ln \left( \frac{\mu^{2}}{m_{Q}^{2}} \right) \right)}{3675 \pi^{4}}
        v^{7}
        +
        \mathcal{O}(v^{8})
    \right]
    \, ,
    \\
    \label{eq:rho-1-th}
    &
    \left. \rho_{1, 1}^{\text{NLO}} \right|_{m_{Q}^{\overline{\text{MS}}}, q^{2} \to 4 m_{Q}^{2}}
    \nonumber \\
    = {} &
    \left. \rho_{1, 1}^{\text{NLO}} \right|_{m_{Q}^{\text{on-shell}}, q^{2} \to 4 m_{Q}^{2}}
    +
    \frac{\alpha_{s}}{2 \pi}
    \left[ \frac{8}{3} + 2 \ln \left( \frac{\mu^{2}}{m_{Q}^{2}} \right) \right]
    (4 m_{Q}^{2})^{2}
    \left[
        -
        \frac{3}{20 \pi^{4}}
        \left( 1 + 4 |\theta|^{2} \right)
        v^{5}
        -
        \frac{3}{140 \pi^{4}}
        \left( 5 + 28 |\theta|^{2} \right)
        v^{7}
        +
        \mathcal{O}(v^{8})
    \right]
    \, ,
    \\
    \label{eq:rho-2-th}
    &
    \left. \rho_{2, 1}^{\text{NLO}} \right|_{m_{Q}^{\overline{\text{MS}}}, q^{2} \to 4 m_{Q}^{2}}
    \nonumber \\
    = {} &
    \left. \rho_{2, 1}^{\text{NLO}} \right|_{m_{Q}^{\text{on-shell}}, q^{2} \to 4 m_{Q}^{2}}
    +
    \frac{\alpha_{s}}{2 \pi}
    \left[ \frac{8}{3} + 2 \ln \left( \frac{\mu^{2}}{m_{Q}^{2}} \right) \right]
    (4 m_{Q}^{2})^{2}
    m_{Q}
    \Im \theta
    \left[
        \frac{6}{5 \pi^{4}}
        v^{5}
        +
        \frac{54}{35 \pi^{4}}
        v^{7}
        +
        \mathcal{O}(v^{8})
    \right]
    \, .
\end{align}
\end{widetext}
The $v^{6}$ terms in above expressions correspond to the Coulombic singularities generated by the gluon exchange between two heavy quarks.
The $v^{5}$ terms in Eq.~(\ref{eq:rho-1-th}) and Eq.~(\ref{eq:rho-2-th}) come from the renormalization scheme difference of $m_{Q}$,
i.e. Eq.~(\ref{eq:diff-1}) and Eq.~(\ref{eq:diff-2}).

We also present our NLO result before renormalization in terms of the coefficients of master integrals
\begin{equation}
    C_{1}^{\text{NLO}}
    =
    i \pi
    \left( \frac{\mu^{2}}{q^{2}} \right)^{3 \varepsilon}
    \sum_{k} (\alpha_{k} \slashed{q} + \beta_{k}) I_{k}
    \, ,
\end{equation}
where $\alpha_{k}$ and $\beta_{k}$ are real.
Thus by definition Eq.~(\ref{eq:rho}),
we have
\begin{align}
    \rho_{1, 1}^{\text{NLO}}
    & =
    \left( \frac{\mu^{2}}{q^{2}} \right)^{3 \varepsilon}
    \sum_{k} \alpha_{k} \Re I_{k}
    \, ,
    \\
    \rho_{2, 1}^{\text{NLO}}
    & =
    \left( \frac{\mu^{2}}{q^{2}} \right)^{3 \varepsilon}
    \sum_{k} \beta_{k} \Re I_{k}
    \, ,
\end{align}
where the master integrals $I_{k}$ are defined to be dimensionless,
which are the same as those in Eq.~(\ref{eq:amp}).
Note that the 29 master integrals in Eq.~(\ref{eq:amp}) contain some symmetries,
that is, some of them can be related to each other by shifting loop momenta.
After using these symmetries, we are only left with 14 master integrals,
which are defined as
\begin{widetext}
\begin{align}
    I_{1}
    & =
    (q^{2})^{3 \varepsilon - 3}
    \prod_{i = 1}^{3}
    \int
    \frac{\mathrm{d}^{d} l_{i}}{(2 \pi)^{d}}
    \frac{1}{l_{i}^{2} - m_{Q}^{2}}
    \, ,
    \\
    I_{2}
    & =
    (q^{2})^{3 \varepsilon - 2}
    \int
    \left( \prod_{i = 1}^{3} \frac{\mathrm{d}^{d} l_{i}}{(2 \pi)^{d}} \right)
    \frac{1}{(l_{1} + l_{2} + l_{3})^{2} l_{1}^{2} [l_{2}^{2} - m_{Q}^{2}] [l_{3}^{2} - m_{Q}^{2}]}
    \, ,
    \\
    I_{3}
    & =
    (q^{2})^{3 \varepsilon - 2}
    \int
    \left( \prod_{i = 1}^{3} \frac{\mathrm{d}^{d} l_{i}}{(2 \pi)^{d}} \right)
    \frac{1}{[(l_{1} + l_{2} + l_{3})^{2} - m_{Q}^{2}] [l_{1}^{2} - m_{Q}^{2}] [l_{2}^{2} - m_{Q}^{2}] [l_{3}^{2} - m_{Q}^{2}]}
    \, ,
    \\
    I_{4}
    & =
    (q^{2})^{3 \varepsilon - 2}
    \int
    \left( \prod_{i = 1}^{3} \frac{\mathrm{d}^{d} l_{i}}{(2 \pi)^{d}} \right)
    \frac{1}{(q - l_{1} - l_{2} - l_{3})^{2} l_{1}^{2} [l_{2}^{2} - m_{Q}^{2}] [l_{3}^{2} - m_{Q}^{2}]}
    \, ,
    \\
    I_{5}
    & =
    (q^{2})^{3 \varepsilon - 1}
    \int
    \left( \prod_{i = 1}^{3} \frac{\mathrm{d}^{d} l_{i}}{(2 \pi)^{d}} \right)
    \frac{1}{(q - l_{1} - l_{2} - l_{3})^{4} l_{1}^{2} [l_{2}^{2} - m_{Q}^{2}] [l_{3}^{2} - m_{Q}^{2}]}
    \, ,
    \\
    I_{6}
    & =
    (q^{2})^{3 \varepsilon - 1}
    \int
    \left( \prod_{i = 1}^{3} \frac{\mathrm{d}^{d} l_{i}}{(2 \pi)^{d}} \right)
    \frac{1}{(q - l_{1} - l_{2} - l_{3})^{2} l_{1}^{2} [l_{2}^{2} - m_{Q}^{2}] [l_{3}^{2} - m_{Q}^{2}]^{2}}
    \, ,
    \\
    I_{7}
    & =
    (q^{2})^{3 \varepsilon - 2}
    \int
    \left( \prod_{i = 1}^{2} \frac{\mathrm{d}^{d} l_{i}}{(2 \pi)^{d}} \right)
    \frac{1}{(q - l_{1} - l_{2})^{2} [l_{1}^{2} - m_{Q}^{2}] [l_{2}^{2} - m_{Q}^{2}]}
    \int
    \frac{\mathrm{d}^{d} l_{3}}{(2 \pi)^{d}}
    \frac{1}{l_{3}^{2} - m_{Q}^{2}}
    \, ,
    \\
    I_{8}
    & =
    (q^{2})^{3 \varepsilon - 1}
    \int
    \left( \prod_{i = 1}^{2} \frac{\mathrm{d}^{d} l_{i}}{(2 \pi)^{d}} \right)
    \frac{1}{(q - l_{1} - l_{2})^{4} [l_{1}^{2} - m_{Q}^{2}] [l_{2}^{2} - m_{Q}^{2}]}
    \int
    \frac{\mathrm{d}^{d} l_{3}}{(2 \pi)^{d}}
    \frac{1}{l_{3}^{2} - m_{Q}^{2}}
    \, ,
    \\
    I_{9}
    & =
    (q^{2})^{3 \varepsilon - 1}
    \int
    \left( \prod_{i = 1}^{2} \frac{\mathrm{d}^{d} l_{i}}{(2 \pi)^{d}} \right)
    \frac{1}{(q - l_{1} - l_{2})^{2} [l_{1}^{2} - m_{Q}^{2}] [l_{2}^{2} - m_{Q}^{2}]^{2}}
    \int
    \frac{\mathrm{d}^{d} l_{3}}{(2 \pi)^{d}}
    \frac{1}{l_{3}^{2} - m_{Q}^{2}}
    \, ,
    \\
    I_{10}
    & =
    (q^{2})^{3 \varepsilon - 1}
    \int
    \left( \prod_{i = 1}^{3} \frac{\mathrm{d}^{d} l_{i}}{(2 \pi)^{d}} \right)
    \frac{1}{(q - l_{1})^{2} [(l_{1} - l_{2})^{2} - m_{Q}^{2}] [l_{2}^{2} - m_{Q}^{2}] [(l_{1} - l_{3})^{2} - m_{Q}^{2}] [l_{3}^{2} - m_{Q}^{2}]}
    \, ,
    \\
    I_{11}
    & =
    (q^{2})^{3 \varepsilon}
    \int
    \left( \prod_{i = 1}^{3} \frac{\mathrm{d}^{d} l_{i}}{(2 \pi)^{d}} \right)
    \frac{1}{(q - l_{1})^{4} [(l_{1} - l_{2})^{2} - m_{Q}^{2}] [l_{2}^{2} - m_{Q}^{2}] [(l_{1} - l_{3})^{2} - m_{Q}^{2}] [l_{3}^{2} - m_{Q}^{2}]}
    \, ,
    \\
    I_{12}
    & =
    (q^{2})^{3 \varepsilon - 1}
    \int
    \left( \prod_{i = 1}^{3} \frac{\mathrm{d}^{d} l_{i}}{(2 \pi)^{d}} \right)
    \frac{1}{[(q - l_{1})^{2} - m_{Q}^{2}] (l_{1} - l_{2})^{2} [l_{2}^{2} - m_{Q}^{2}] (l_{1} - l_{3})^{2} [l_{3}^{2} - m_{Q}^{2}]}
    \, ,
    \\
    I_{13}
    & =
    (q^{2})^{3 \varepsilon}
    \int
    \left( \prod_{i = 1}^{3} \frac{\mathrm{d}^{d} l_{i}}{(2 \pi)^{d}} \right)
    \frac{1}{[(q - l_{1})^{2} - m_{Q}^{2}]^{2} (l_{1} - l_{2})^{2} [l_{2}^{2} - m_{Q}^{2}] (l_{1} - l_{3})^{2} [l_{3}^{2} - m_{Q}^{2}]}
    \, ,
    \\
    I_{14}
    & =
    (q^{2})^{3 \varepsilon}
    \int
    \left( \prod_{i = 1}^{3} \frac{\mathrm{d}^{d} l_{i}}{(2 \pi)^{d}} \right)
    \frac{1}{[(q - l_{1})^{2} - m_{Q}^{2}] (l_{1} - l_{2})^{2} [l_{2}^{2} - m_{Q}^{2}] (l_{1} - l_{3})^{2} [l_{3}^{2} - m_{Q}^{2}]^{2}}
    \, .
\end{align}
\end{widetext}
Using the differential equation method, we obtain the real part of master integrals up to $\mathcal{O}(\varepsilon^{2})$ in terms of Goncharov polylogarithms.
The explicit expressions of $\alpha_{i}$, $\beta_{i}$ and $\Re I_{i}$ are lengthy and will be presented in the ancillary file of the arXiv preprint.

\sect{Higher Dimensional Operators}
\label{sec:ope}
In addition to $\langle \overline{q} q \rangle$ and $\langle g_{s}^{2} G G \rangle$ operators,
we also calculate the Wilson coefficients of $\langle g_{s} \overline{q} q G \rangle$ operator up to the leading contributions
\begin{align}
    \rho_{1, \langle g_{s} \overline{q} q G \rangle}^{\text{LO}}
    & =
    \rho_{c}^{\langle g_{s} \overline{q} q G \rangle}
    +
    \rho_{c}^{\langle \overline{q} q \rangle}
    \, ,
    \\
    \rho_{2, \langle g_{s} \overline{q} q G \rangle}^{\text{LO}}
    & =
    \rho_{d}^{\langle g_{s} \overline{q} q G \rangle}
    +
    \rho_{d}^{\langle \overline{q} q \rangle}
    \, ,
\end{align}
where $\rho_{c}^{\langle g_{s} \overline{q} q G \rangle}$ and $\rho_{d}^{\langle g_{s} \overline{q} q G \rangle}$ come directly from the $\langle g_{s} \overline{q} q G \rangle$ operator,
while $\rho_{c}^{\langle \overline{q} q \rangle}$ and $\rho_{d}^{\langle \overline{q} q \rangle}$ are contributions from the expansion of $\langle \overline{q} q \rangle$ operator \cite{Ioffe:1981kw}.
Here $\rho_{c}^{\langle g_{s} \overline{q} q G \rangle}$ and $\rho_{d}^{\langle g_{s} \overline{q} q G \rangle}$ are
\begin{align}
    \rho_{c}^{\langle g_{s} \overline{q} q G \rangle}
    & =
    \frac{1}{8 \pi^{2}} \frac{m_{Q}}{q^{2}} \Im \theta \frac{5 v^{2} + 7}{v}
    \, ,
    \\
    \rho_{d}^{\langle g_{s} \overline{q} q G \rangle}
    & =
    - \frac{1}{32 \pi^{2}} \frac{v^{2} + 3}{v}
    +
    \frac{1}{2 \pi^{2}} |\theta|^{2} \frac{v^{2} - 1}{v}
    \, ,
\end{align}
and $\rho_{c}^{\langle \overline{q} q \rangle}$ and $\rho_{d}^{\langle \overline{q} q \rangle}$ are
\begin{align}
    \rho_{c}^{\langle \overline{q} q \rangle}
    & =
    \frac{3}{16 \pi^{2}} \frac{m_{Q}}{q^{2}} \Im \theta \frac{(v^{2} - 1) (3 v^{2} - 1)}{v^{3}}
    \, ,
    \\
    \rho_{d}^{\langle \overline{q} q \rangle}
    & =
    \frac{1}{128 \pi^{2}} \frac{v^{6} + 3 v^{4} + 15 v^{2} - 3}{v^{3}}
    +
    \frac{3}{32 \pi^{2}} |\theta|^{2} \frac{(v^{2} - 1)^{3}}{v^{3}}
    \, .
\end{align}
Again, with the help of Eq.~(\ref{eq:mix}),
our result confirms previous calculations \cite{Bagan:1992za}.

Note that $\rho_{c}^{\langle \overline{q} q \rangle}$ and $\rho_{d}^{\langle \overline{q} q \rangle}$ contain Coulombic-like singularities,
which will cause the integral over $s$ in Eq.~(\ref{eq:sum-1}) to diverge at the threshold.
Thus we cannot use the above results in our sum rule analysis directly.
To deal with these singularities,
we may consider resumming the leading Coulombic interaction $\left( \frac{\alpha_{s}}{v} \right)^{n}$ between two heavy quark $Q$.
The amplitude of $(Q^{a} C \Gamma_{1} Q^{b})$ part of the baryon current is multiplied by the Sommerfeld factor
\cite{Sommerfeld:1741612, Kiselev:1999zj, Portoles:2002rt}
\begin{equation}
    S(v)
    =
    \frac{\frac{C \pi \alpha_{s}}{v}}{1 - \exp \left( - \frac{C \pi \alpha_{s}}{v} \right)}
    \, ,
\end{equation}
where $C$ is the color factor.
In our case,
$Q^{a} Q^{b}$ forms a color anti-triplet, so we have $C = \frac{2}{3}$.
Then we calculate Wilson coefficients of the $\langle g_{s} \overline{q} q G \rangle$ operator as before.
The resummed $\rho_{c}^{\langle \overline{q} q \rangle}$ and $\rho_{d}^{\langle \overline{q} q \rangle}$ are
\begin{widetext}
\begin{align}
    \rho_{c}^{\langle \overline{q} q \rangle}
    & =
    \frac{3}{16 \pi^{2}} \frac{m_{Q}}{q^{2}} \Im \theta \frac{(v^{2} - 1)}{v^{3}} \left[
        S(v) (3 v^{2} - 1)
        +
        S'(v) v (v^{2} + 1)
        -
        S''(v) v^{2} (v^{2} - 1)
    \right]
    \, ,
    \\
    \rho_{d}^{\langle \overline{q} q \rangle}
    & =
    \frac{1}{128 \pi^{2}} \frac{1}{v^{3}} \left[
        S(v) (v^{6} + 3 v^{4} + 15 v^{2} - 3)
        -
        S'(v) v (v^{4} - 1) (v^{2} + 3)
        -
        S''(v) v^{2} (v^{2} - 1)^{2} (v^{2} - 3)
    \right]
    \nonumber \\
    & \phantom{= {}} +
    \frac{3}{32 \pi^{2}} |\theta|^{2} \frac{(v^{2} - 1)^{3}}{v^{3}} \left[
        S(v) - S'(v) v - S''(v) v^{2}
    \right]
    \, .
\end{align}
\end{widetext}
After resummation,
the Coulombic-like singularities are regularized by the Sommerfeld factor,
and the integral over $s$ in Eq.~(\ref{eq:sum-1}) converges.

\begin{figure}[ht]
    \centering
    \includegraphics[width=0.6\linewidth]{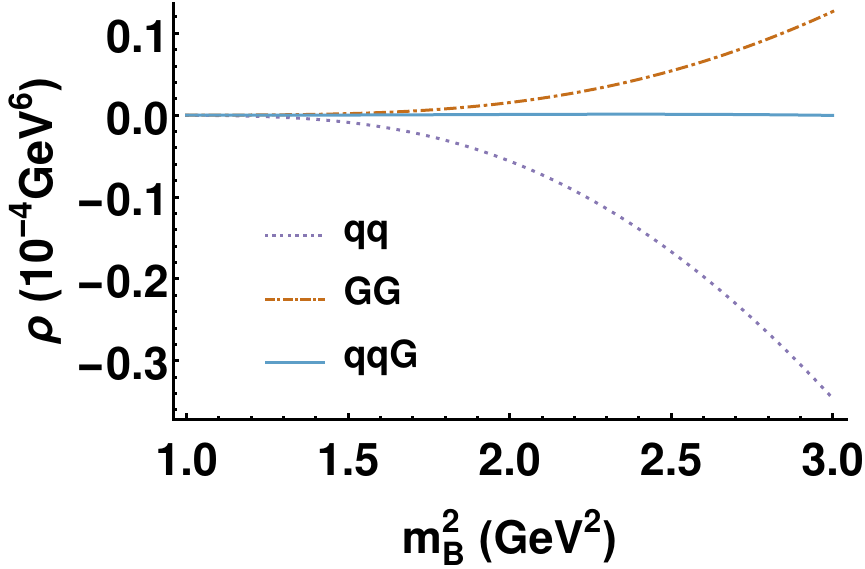}
    \caption{
        Contributions of various terms on the right hand side of Eq.~(\ref{eq:sum-1}).
    }
    \label{fig:qqg-ope}
\end{figure}

Now we can include the $\langle g_{s} \overline{q} q G \rangle$ condensate in our sum rule analysis,
and investigate its contribution to the sum rule and $m_{\Xi_{cc}^{++}}$ estimation.
The vacuum condensate parameter is taken to be \cite{Bagan:1992za, Kiselev:1999zj, Zhang:2008rt, Wang:2010hs, Narison:2010py, Tang:2011fv, Aliev:2012ru}
\begin{equation}
    \langle g_{s} \overline{q} q G \rangle(2 \text{~GeV})
    =
    (0.8 \pm 0.2 \text{~GeV}^{2})
    \times
    \langle \overline{q} q \rangle(2 \text{~GeV})
    \, .
\end{equation}
The vacuum condensate can be evolved according to its one-loop anomalous dimensions:
$\gamma_{\langle g_{s} \overline{q} q G \rangle} = - \frac{\gamma_{m_{q}}}{6}$ \cite{Albuquerque:2013ija}.
The relative importance of each condensate term in OPE, including $\langle g_{s} \overline{q} q G \rangle$, is shown in Fig.~(\ref{fig:qqg-ope}).

Define the condensate term of $O_{i}$ to be
\begin{equation}
    c_{i}
    =
    \langle O_{i} \rangle
    \int_{s_{th}}^{\infty} \mathrm{d} s \,
    \rho_{1, i}(s)
    e^{- \frac{s}{m_{B}^{2}}}
    \, ,
\end{equation}
the ratios between consecutive terms in the $m_{Q}^{\overline{\text{MS}}}$ scheme at central values of all parameters are
\begin{equation}
    \left| \frac{c_{g_{s}^{2} G G}}{c_{\overline{q} q}} \right|
    =
    24 \%
    \, ,
    \quad
    \left| \frac{c_{g_{s} \overline{q} q G}}{c_{g_{s}^{2} G G}} \right|
    =
    8 \%
    \, .
\end{equation}
$c_{g_{s} \overline{q} q G}$ contribution to the right hand side of Eq.~(\ref{eq:sum-1}) is less than $0.6 \%$,
and the estimated $m_{\Xi_{cc}^{++}}$ changes by less than $0.3 \%$ in both LO and $\text{LO} + \text{NLO}$ cases.
We see that the OPE seems to show good convergence,
and it might be a good approximation to neglect the contributions of operators with dimension larger than 4 in the sum rule Eq.~(\ref{eq:sum-1}).
Nevertheless, it is certainly helpful to have a systematical study for the contributions of higher dimensional operators in the future.


\begin{thebibliography}{58}%
\makeatletter
\providecommand \@ifxundefined [1]{%
 \@ifx{#1\undefined}
}%
\providecommand \@ifnum [1]{%
 \ifnum #1\expandafter \@firstoftwo
 \else \expandafter \@secondoftwo
 \fi
}%
\providecommand \@ifx [1]{%
 \ifx #1\expandafter \@firstoftwo
 \else \expandafter \@secondoftwo
 \fi
}%
\providecommand \natexlab [1]{#1}%
\providecommand \enquote  [1]{``#1''}%
\providecommand \bibnamefont  [1]{#1}%
\providecommand \bibfnamefont [1]{#1}%
\providecommand \citenamefont [1]{#1}%
\providecommand \href@noop [0]{\@secondoftwo}%
\providecommand \href [0]{\begingroup \@sanitize@url \@href}%
\providecommand \@href[1]{\@@startlink{#1}\@@href}%
\providecommand \@@href[1]{\endgroup#1\@@endlink}%
\providecommand \@sanitize@url [0]{\catcode `\\12\catcode `\$12\catcode
  `\&12\catcode `\#12\catcode `\^12\catcode `\_12\catcode `\%12\relax}%
\providecommand \@@startlink[1]{}%
\providecommand \@@endlink[0]{}%
\providecommand \url  [0]{\begingroup\@sanitize@url \@url }%
\providecommand \@url [1]{\endgroup\@href {#1}{\urlprefix }}%
\providecommand \urlprefix  [0]{URL }%
\providecommand \Eprint [0]{\href }%
\providecommand \doibase [0]{http://dx.doi.org/}%
\providecommand \selectlanguage [0]{\@gobble}%
\providecommand \bibinfo  [0]{\@secondoftwo}%
\providecommand \bibfield  [0]{\@secondoftwo}%
\providecommand \translation [1]{[#1]}%
\providecommand \BibitemOpen [0]{}%
\providecommand \bibitemStop [0]{}%
\providecommand \bibitemNoStop [0]{.\EOS\space}%
\providecommand \EOS [0]{\spacefactor3000\relax}%
\providecommand \BibitemShut  [1]{\csname bibitem#1\endcsname}%
\let\auto@bib@innerbib\@empty
\bibitem [{\citenamefont {Aaij}\ \emph {et~al.}(2017)\citenamefont {Aaij} \emph
  {et~al.}}]{Aaij:2017ueg}%
  \BibitemOpen
  \bibfield  {author} {\bibinfo {author} {\bibfnamefont {R.}~\bibnamefont
  {Aaij}} \emph {et~al.} (\bibinfo {collaboration} {LHCb}),\ }\href@noop {} {\
  (\bibinfo {year} {2017})},\ \Eprint {http://arxiv.org/abs/1707.01621}
  {arXiv:1707.01621 [hep-ex]} \BibitemShut {NoStop}%
\bibitem [{\citenamefont {Mattson}\ \emph {et~al.}(2002)\citenamefont {Mattson}
  \emph {et~al.}}]{Mattson:2002vu}%
  \BibitemOpen
  \bibfield  {author} {\bibinfo {author} {\bibfnamefont {M.}~\bibnamefont
  {Mattson}} \emph {et~al.} (\bibinfo {collaboration} {SELEX}),\ }\href
  {\doibase 10.1103/PhysRevLett.89.112001} {\bibfield  {journal} {\bibinfo
  {journal} {Phys. Rev. Lett.}\ }\textbf {\bibinfo {volume} {89}},\ \bibinfo
  {pages} {112001} (\bibinfo {year} {2002})},\ \Eprint
  {http://arxiv.org/abs/hep-ex/0208014} {arXiv:hep-ex/0208014 [hep-ex]}
  \BibitemShut {NoStop}%
\bibitem [{\citenamefont {Aubert}\ \emph {et~al.}(2006)\citenamefont {Aubert}
  \emph {et~al.}}]{Aubert:2006qw}%
  \BibitemOpen
  \bibfield  {author} {\bibinfo {author} {\bibfnamefont {B.}~\bibnamefont
  {Aubert}} \emph {et~al.} (\bibinfo {collaboration} {BaBar}),\ }\href
  {\doibase 10.1103/PhysRevD.74.011103} {\bibfield  {journal} {\bibinfo
  {journal} {Phys. Rev.}\ }\textbf {\bibinfo {volume} {D74}},\ \bibinfo {pages}
  {011103} (\bibinfo {year} {2006})},\ \Eprint
  {http://arxiv.org/abs/hep-ex/0605075} {arXiv:hep-ex/0605075 [hep-ex]}
  \BibitemShut {NoStop}%
\bibitem [{\citenamefont {Chistov}\ \emph {et~al.}(2006)\citenamefont {Chistov}
  \emph {et~al.}}]{Chistov:2006zj}%
  \BibitemOpen
  \bibfield  {author} {\bibinfo {author} {\bibfnamefont {R.}~\bibnamefont
  {Chistov}} \emph {et~al.} (\bibinfo {collaboration} {Belle}),\ }\href
  {\doibase 10.1103/PhysRevLett.97.162001} {\bibfield  {journal} {\bibinfo
  {journal} {Phys. Rev. Lett.}\ }\textbf {\bibinfo {volume} {97}},\ \bibinfo
  {pages} {162001} (\bibinfo {year} {2006})},\ \Eprint
  {http://arxiv.org/abs/hep-ex/0606051} {arXiv:hep-ex/0606051 [hep-ex]}
  \BibitemShut {NoStop}%
\bibitem [{\citenamefont {Hudspith}\ \emph {et~al.}(2017)\citenamefont
  {Hudspith}, \citenamefont {Francis}, \citenamefont {Lewis},\ and\
  \citenamefont {Maltman}}]{Hudspith:2017bbh}%
  \BibitemOpen
  \bibfield  {author} {\bibinfo {author} {\bibfnamefont {R.~J.}\ \bibnamefont
  {Hudspith}}, \bibinfo {author} {\bibfnamefont {A.}~\bibnamefont {Francis}},
  \bibinfo {author} {\bibfnamefont {R.}~\bibnamefont {Lewis}}, \ and\ \bibinfo
  {author} {\bibfnamefont {K.}~\bibnamefont {Maltman}},\ }\bibfield
  {booktitle} {\emph {\bibinfo {booktitle} {{Proceedings, 34th International
  Symposium on Lattice Field Theory (Lattice 2016): Southampton, UK, July
  24-30, 2016}}},\ }\href@noop {} {\bibfield  {journal} {\bibinfo  {journal}
  {PoS}\ }\textbf {\bibinfo {volume} {LATTICE2016}},\ \bibinfo {pages} {133}
  (\bibinfo {year} {2017})}\BibitemShut {NoStop}%
\bibitem [{\citenamefont {Namekawa}\ \emph {et~al.}(2013)\citenamefont
  {Namekawa} \emph {et~al.}}]{Namekawa:2013vu}%
  \BibitemOpen
  \bibfield  {author} {\bibinfo {author} {\bibfnamefont {Y.}~\bibnamefont
  {Namekawa}} \emph {et~al.} (\bibinfo {collaboration} {PACS-CS}),\ }\href
  {\doibase 10.1103/PhysRevD.87.094512} {\bibfield  {journal} {\bibinfo
  {journal} {Phys. Rev.}\ }\textbf {\bibinfo {volume} {D87}},\ \bibinfo {pages}
  {094512} (\bibinfo {year} {2013})},\ \Eprint {http://arxiv.org/abs/1301.4743}
  {arXiv:1301.4743 [hep-lat]} \BibitemShut {NoStop}%
\bibitem [{\citenamefont {Lewis}\ \emph {et~al.}(2001)\citenamefont {Lewis},
  \citenamefont {Mathur},\ and\ \citenamefont {Woloshyn}}]{Lewis:2001iz}%
  \BibitemOpen
  \bibfield  {author} {\bibinfo {author} {\bibfnamefont {R.}~\bibnamefont
  {Lewis}}, \bibinfo {author} {\bibfnamefont {N.}~\bibnamefont {Mathur}}, \
  and\ \bibinfo {author} {\bibfnamefont {R.~M.}\ \bibnamefont {Woloshyn}},\
  }\href {\doibase 10.1103/PhysRevD.64.094509} {\bibfield  {journal} {\bibinfo
  {journal} {Phys. Rev.}\ }\textbf {\bibinfo {volume} {D64}},\ \bibinfo {pages}
  {094509} (\bibinfo {year} {2001})},\ \Eprint
  {http://arxiv.org/abs/hep-ph/0107037} {arXiv:hep-ph/0107037 [hep-ph]}
  \BibitemShut {NoStop}%
\bibitem [{\citenamefont {Sun}\ and\ \citenamefont
  {Vicente~Vacas}(2016)}]{Sun:2016wzh}%
  \BibitemOpen
  \bibfield  {author} {\bibinfo {author} {\bibfnamefont {Z.-F.}\ \bibnamefont
  {Sun}}\ and\ \bibinfo {author} {\bibfnamefont {M.~J.}\ \bibnamefont
  {Vicente~Vacas}},\ }\href {\doibase 10.1103/PhysRevD.93.094002} {\bibfield
  {journal} {\bibinfo  {journal} {Phys. Rev.}\ }\textbf {\bibinfo {volume}
  {D93}},\ \bibinfo {pages} {094002} (\bibinfo {year} {2016})},\ \Eprint
  {http://arxiv.org/abs/1602.04714} {arXiv:1602.04714 [hep-ph]} \BibitemShut
  {NoStop}%
\bibitem [{\citenamefont {Kiselev}\ \emph {et~al.}(2017)\citenamefont
  {Kiselev}, \citenamefont {Berezhnoy},\ and\ \citenamefont
  {Likhoded}}]{Kiselev:2017eic}%
  \BibitemOpen
  \bibfield  {author} {\bibinfo {author} {\bibfnamefont {A.~V.}\ \bibnamefont
  {Kiselev}}, \bibinfo {author} {\bibfnamefont {A.~V.}\ \bibnamefont
  {Berezhnoy}}, \ and\ \bibinfo {author} {\bibfnamefont {A.~K.}\ \bibnamefont
  {Likhoded}},\ }\href@noop {} {\  (\bibinfo {year} {2017})},\ \Eprint
  {http://arxiv.org/abs/1706.09181} {arXiv:1706.09181 [hep-ph]} \BibitemShut
  {NoStop}%
\bibitem [{\citenamefont {Shah}\ and\ \citenamefont
  {Rai}(2017)}]{Shah:2017liu}%
  \BibitemOpen
  \bibfield  {author} {\bibinfo {author} {\bibfnamefont {Z.}~\bibnamefont
  {Shah}}\ and\ \bibinfo {author} {\bibfnamefont {A.~K.}\ \bibnamefont {Rai}},\
  }\href {\doibase 10.1140/epjc/s10052-017-4688-x} {\bibfield  {journal}
  {\bibinfo  {journal} {Eur. Phys. J.}\ }\textbf {\bibinfo {volume} {C77}},\
  \bibinfo {pages} {129} (\bibinfo {year} {2017})},\ \Eprint
  {http://arxiv.org/abs/1702.02726} {arXiv:1702.02726 [hep-ph]} \BibitemShut
  {NoStop}%
\bibitem [{\citenamefont {Gadaria}\ \emph {et~al.}(2016)\citenamefont
  {Gadaria}, \citenamefont {Soni},\ and\ \citenamefont
  {Pandya}}]{Gadaria:2016omw}%
  \BibitemOpen
  \bibfield  {author} {\bibinfo {author} {\bibfnamefont {A.~N.}\ \bibnamefont
  {Gadaria}}, \bibinfo {author} {\bibfnamefont {N.~R.}\ \bibnamefont {Soni}}, \
  and\ \bibinfo {author} {\bibfnamefont {J.~N.}\ \bibnamefont {Pandya}},\
  }\bibfield  {booktitle} {\emph {\bibinfo {booktitle} {{Proceedings, 61st
  DAE-BRNS Symposium on Nuclear Physics: Kolkata, India, 5-9 December 2016}}},\
  }\href@noop {} {\bibfield  {journal} {\bibinfo  {journal} {DAE Symp. Nucl.
  Phys.}\ }\textbf {\bibinfo {volume} {61}},\ \bibinfo {pages} {698} (\bibinfo
  {year} {2016})}\BibitemShut {NoStop}%
\bibitem [{\citenamefont {Roberts}\ and\ \citenamefont
  {Pervin}(2008)}]{Roberts:2007ni}%
  \BibitemOpen
  \bibfield  {author} {\bibinfo {author} {\bibfnamefont {W.}~\bibnamefont
  {Roberts}}\ and\ \bibinfo {author} {\bibfnamefont {M.}~\bibnamefont
  {Pervin}},\ }\href {\doibase 10.1142/S0217751X08041219} {\bibfield  {journal}
  {\bibinfo  {journal} {Int. J. Mod. Phys.}\ }\textbf {\bibinfo {volume}
  {A23}},\ \bibinfo {pages} {2817} (\bibinfo {year} {2008})},\ \Eprint
  {http://arxiv.org/abs/0711.2492} {arXiv:0711.2492 [nucl-th]} \BibitemShut
  {NoStop}%
\bibitem [{\citenamefont {Ebert}\ \emph {et~al.}(2002)\citenamefont {Ebert},
  \citenamefont {Faustov}, \citenamefont {Galkin},\ and\ \citenamefont
  {Martynenko}}]{Ebert:2002ig}%
  \BibitemOpen
  \bibfield  {author} {\bibinfo {author} {\bibfnamefont {D.}~\bibnamefont
  {Ebert}}, \bibinfo {author} {\bibfnamefont {R.~N.}\ \bibnamefont {Faustov}},
  \bibinfo {author} {\bibfnamefont {V.~O.}\ \bibnamefont {Galkin}}, \ and\
  \bibinfo {author} {\bibfnamefont {A.~P.}\ \bibnamefont {Martynenko}},\ }\href
  {\doibase 10.1103/PhysRevD.66.014008} {\bibfield  {journal} {\bibinfo
  {journal} {Phys. Rev.}\ }\textbf {\bibinfo {volume} {D66}},\ \bibinfo {pages}
  {014008} (\bibinfo {year} {2002})},\ \Eprint
  {http://arxiv.org/abs/hep-ph/0201217} {arXiv:hep-ph/0201217 [hep-ph]}
  \BibitemShut {NoStop}%
\bibitem [{\citenamefont {Shifman}\ \emph
  {et~al.}(1979{\natexlab{a}})\citenamefont {Shifman}, \citenamefont
  {Vainshtein},\ and\ \citenamefont {Zakharov}}]{Shifman:1978bx}%
  \BibitemOpen
  \bibfield  {author} {\bibinfo {author} {\bibfnamefont {M.~A.}\ \bibnamefont
  {Shifman}}, \bibinfo {author} {\bibfnamefont {A.~I.}\ \bibnamefont
  {Vainshtein}}, \ and\ \bibinfo {author} {\bibfnamefont {V.~I.}\ \bibnamefont
  {Zakharov}},\ }\href {\doibase 10.1016/0550-3213(79)90022-1} {\bibfield
  {journal} {\bibinfo  {journal} {Nucl. Phys.}\ }\textbf {\bibinfo {volume}
  {B147}},\ \bibinfo {pages} {385} (\bibinfo {year}
  {1979}{\natexlab{a}})}\BibitemShut {NoStop}%
\bibitem [{\citenamefont {Shifman}\ \emph
  {et~al.}(1979{\natexlab{b}})\citenamefont {Shifman}, \citenamefont
  {Vainshtein},\ and\ \citenamefont {Zakharov}}]{Shifman:1978by}%
  \BibitemOpen
  \bibfield  {author} {\bibinfo {author} {\bibfnamefont {M.~A.}\ \bibnamefont
  {Shifman}}, \bibinfo {author} {\bibfnamefont {A.~I.}\ \bibnamefont
  {Vainshtein}}, \ and\ \bibinfo {author} {\bibfnamefont {V.~I.}\ \bibnamefont
  {Zakharov}},\ }\href {\doibase 10.1016/0550-3213(79)90023-3} {\bibfield
  {journal} {\bibinfo  {journal} {Nucl. Phys.}\ }\textbf {\bibinfo {volume}
  {B147}},\ \bibinfo {pages} {448} (\bibinfo {year}
  {1979}{\natexlab{b}})}\BibitemShut {NoStop}%
\bibitem [{\citenamefont {Bagan}\ \emph {et~al.}(1993)\citenamefont {Bagan},
  \citenamefont {Chabab},\ and\ \citenamefont {Narison}}]{Bagan:1992za}%
  \BibitemOpen
  \bibfield  {author} {\bibinfo {author} {\bibfnamefont {E.}~\bibnamefont
  {Bagan}}, \bibinfo {author} {\bibfnamefont {M.}~\bibnamefont {Chabab}}, \
  and\ \bibinfo {author} {\bibfnamefont {S.}~\bibnamefont {Narison}},\ }\href
  {\doibase 10.1016/0370-2693(93)90090-5} {\bibfield  {journal} {\bibinfo
  {journal} {Phys. Lett.}\ }\textbf {\bibinfo {volume} {B306}},\ \bibinfo
  {pages} {350} (\bibinfo {year} {1993})}\BibitemShut {NoStop}%
\bibitem [{\citenamefont {Kiselev}\ and\ \citenamefont
  {Onishchenko}(2000)}]{Kiselev:1999zj}%
  \BibitemOpen
  \bibfield  {author} {\bibinfo {author} {\bibfnamefont {V.~V.}\ \bibnamefont
  {Kiselev}}\ and\ \bibinfo {author} {\bibfnamefont {A.~I.}\ \bibnamefont
  {Onishchenko}},\ }\href {\doibase 10.1016/S0550-3213(00)00128-0} {\bibfield
  {journal} {\bibinfo  {journal} {Nucl. Phys.}\ }\textbf {\bibinfo {volume}
  {B581}},\ \bibinfo {pages} {432} (\bibinfo {year} {2000})},\ \Eprint
  {http://arxiv.org/abs/hep-ph/9909337} {arXiv:hep-ph/9909337 [hep-ph]}
  \BibitemShut {NoStop}%
\bibitem [{\citenamefont {Zhang}\ and\ \citenamefont
  {Huang}(2008)}]{Zhang:2008rt}%
  \BibitemOpen
  \bibfield  {author} {\bibinfo {author} {\bibfnamefont {J.-R.}\ \bibnamefont
  {Zhang}}\ and\ \bibinfo {author} {\bibfnamefont {M.-Q.}\ \bibnamefont
  {Huang}},\ }\href {\doibase 10.1103/PhysRevD.78.094007} {\bibfield  {journal}
  {\bibinfo  {journal} {Phys. Rev.}\ }\textbf {\bibinfo {volume} {D78}},\
  \bibinfo {pages} {094007} (\bibinfo {year} {2008})},\ \Eprint
  {http://arxiv.org/abs/0810.5396} {arXiv:0810.5396 [hep-ph]} \BibitemShut
  {NoStop}%
\bibitem [{\citenamefont {Wang}(2010)}]{Wang:2010hs}%
  \BibitemOpen
  \bibfield  {author} {\bibinfo {author} {\bibfnamefont {Z.-G.}\ \bibnamefont
  {Wang}},\ }\href {\doibase 10.1140/epja/i2010-11004-3} {\bibfield  {journal}
  {\bibinfo  {journal} {Eur. Phys. J.}\ }\textbf {\bibinfo {volume} {A45}},\
  \bibinfo {pages} {267} (\bibinfo {year} {2010})},\ \Eprint
  {http://arxiv.org/abs/1001.4693} {arXiv:1001.4693 [hep-ph]} \BibitemShut
  {NoStop}%
\bibitem [{\citenamefont {Tang}\ \emph {et~al.}(2012)\citenamefont {Tang},
  \citenamefont {Yuan}, \citenamefont {Qiao},\ and\ \citenamefont
  {Li}}]{Tang:2011fv}%
  \BibitemOpen
  \bibfield  {author} {\bibinfo {author} {\bibfnamefont {L.}~\bibnamefont
  {Tang}}, \bibinfo {author} {\bibfnamefont {X.-H.}\ \bibnamefont {Yuan}},
  \bibinfo {author} {\bibfnamefont {C.-F.}\ \bibnamefont {Qiao}}, \ and\
  \bibinfo {author} {\bibfnamefont {X.-Q.}\ \bibnamefont {Li}},\ }\href
  {\doibase 10.1088/0253-6102/57/3/15} {\bibfield  {journal} {\bibinfo
  {journal} {Commun. Theor. Phys.}\ }\textbf {\bibinfo {volume} {57}},\
  \bibinfo {pages} {435} (\bibinfo {year} {2012})},\ \Eprint
  {http://arxiv.org/abs/1104.4934} {arXiv:1104.4934 [hep-ph]} \BibitemShut
  {NoStop}%
\bibitem [{\citenamefont {Aliev}\ \emph {et~al.}(2012)\citenamefont {Aliev},
  \citenamefont {Azizi},\ and\ \citenamefont {Savci}}]{Aliev:2012ru}%
  \BibitemOpen
  \bibfield  {author} {\bibinfo {author} {\bibfnamefont {T.~M.}\ \bibnamefont
  {Aliev}}, \bibinfo {author} {\bibfnamefont {K.}~\bibnamefont {Azizi}}, \ and\
  \bibinfo {author} {\bibfnamefont {M.}~\bibnamefont {Savci}},\ }\href
  {\doibase 10.1016/j.nuclphysa.2012.09.009} {\bibfield  {journal} {\bibinfo
  {journal} {Nucl. Phys.}\ }\textbf {\bibinfo {volume} {A895}},\ \bibinfo
  {pages} {59} (\bibinfo {year} {2012})},\ \Eprint
  {http://arxiv.org/abs/1205.2873} {arXiv:1205.2873 [hep-ph]} \BibitemShut
  {NoStop}%
\bibitem [{\citenamefont {Chen}\ \emph {et~al.}(2017)\citenamefont {Chen},
  \citenamefont {Mao}, \citenamefont {Chen}, \citenamefont {Liu},\ and\
  \citenamefont {Zhu}}]{Chen:2017sbg}%
  \BibitemOpen
  \bibfield  {author} {\bibinfo {author} {\bibfnamefont {H.-X.}\ \bibnamefont
  {Chen}}, \bibinfo {author} {\bibfnamefont {Q.}~\bibnamefont {Mao}}, \bibinfo
  {author} {\bibfnamefont {W.}~\bibnamefont {Chen}}, \bibinfo {author}
  {\bibfnamefont {X.}~\bibnamefont {Liu}}, \ and\ \bibinfo {author}
  {\bibfnamefont {S.-L.}\ \bibnamefont {Zhu}},\ }\href@noop {} {\  (\bibinfo
  {year} {2017})},\ \Eprint {http://arxiv.org/abs/1707.01779} {arXiv:1707.01779
  [hep-ph]} \BibitemShut {NoStop}%
\bibitem [{\citenamefont {Reinders}\ \emph {et~al.}(1985)\citenamefont
  {Reinders}, \citenamefont {Rubinstein},\ and\ \citenamefont
  {Yazaki}}]{Reinders:1984sr}%
  \BibitemOpen
  \bibfield  {author} {\bibinfo {author} {\bibfnamefont {L.~J.}\ \bibnamefont
  {Reinders}}, \bibinfo {author} {\bibfnamefont {H.}~\bibnamefont
  {Rubinstein}}, \ and\ \bibinfo {author} {\bibfnamefont {S.}~\bibnamefont
  {Yazaki}},\ }\href {\doibase 10.1016/0370-1573(85)90065-1} {\bibfield
  {journal} {\bibinfo  {journal} {Phys. Rept.}\ }\textbf {\bibinfo {volume}
  {127}},\ \bibinfo {pages} {1} (\bibinfo {year} {1985})}\BibitemShut {NoStop}%
\bibitem [{\citenamefont {Jamin}(1988)}]{Jamin:1987gq}%
  \BibitemOpen
  \bibfield  {author} {\bibinfo {author} {\bibfnamefont {M.}~\bibnamefont
  {Jamin}},\ }\href {\doibase 10.1007/BF01549725} {\bibfield  {journal}
  {\bibinfo  {journal} {Z. Phys.}\ }\textbf {\bibinfo {volume} {C37}},\
  \bibinfo {pages} {635} (\bibinfo {year} {1988})}\BibitemShut {NoStop}%
\bibitem [{\citenamefont {Ovchinnikov}\ \emph {et~al.}(1991)\citenamefont
  {Ovchinnikov}, \citenamefont {Pivovarov},\ and\ \citenamefont
  {Surguladze}}]{Ovchinnikov:1991mu}%
  \BibitemOpen
  \bibfield  {author} {\bibinfo {author} {\bibfnamefont {A.~A.}\ \bibnamefont
  {Ovchinnikov}}, \bibinfo {author} {\bibfnamefont {A.~A.}\ \bibnamefont
  {Pivovarov}}, \ and\ \bibinfo {author} {\bibfnamefont {L.~R.}\ \bibnamefont
  {Surguladze}},\ }\href {\doibase 10.1142/S0217751X91001015} {\bibfield
  {journal} {\bibinfo  {journal} {Int. J. Mod. Phys.}\ }\textbf {\bibinfo
  {volume} {A6}},\ \bibinfo {pages} {2025} (\bibinfo {year}
  {1991})}\BibitemShut {NoStop}%
\bibitem [{\citenamefont {Schwinger}(1989)}]{Schwinger:1989ix}%
  \BibitemOpen
  \bibfield  {author} {\bibinfo {author} {\bibfnamefont {J.~S.}\ \bibnamefont
  {Schwinger}},\ }\href@noop {} {\emph {\bibinfo {title} {{PARTICLES, SOURCES,
  AND FIELDS. VOL. 3}}}}\ (\bibinfo {year} {1989})\BibitemShut {NoStop}%
\bibitem [{\citenamefont {Maier}\ and\ \citenamefont
  {Marquard}(2012)}]{Maier:2011jd}%
  \BibitemOpen
  \bibfield  {author} {\bibinfo {author} {\bibfnamefont {A.}~\bibnamefont
  {Maier}}\ and\ \bibinfo {author} {\bibfnamefont {P.}~\bibnamefont
  {Marquard}},\ }\href {\doibase 10.1016/j.nuclphysb.2012.01.021} {\bibfield
  {journal} {\bibinfo  {journal} {Nucl. Phys.}\ }\textbf {\bibinfo {volume}
  {B859}},\ \bibinfo {pages} {1} (\bibinfo {year} {2012})},\ \Eprint
  {http://arxiv.org/abs/1110.5581} {arXiv:1110.5581 [hep-ph]} \BibitemShut
  {NoStop}%
\bibitem [{\citenamefont {Baikov}\ \emph {et~al.}(2009)\citenamefont {Baikov},
  \citenamefont {Chetyrkin},\ and\ \citenamefont {Kuhn}}]{Baikov:2009uw}%
  \BibitemOpen
  \bibfield  {author} {\bibinfo {author} {\bibfnamefont {P.~A.}\ \bibnamefont
  {Baikov}}, \bibinfo {author} {\bibfnamefont {K.~G.}\ \bibnamefont
  {Chetyrkin}}, \ and\ \bibinfo {author} {\bibfnamefont {J.~H.}\ \bibnamefont
  {Kuhn}},\ }\bibfield  {booktitle} {\emph {\bibinfo {booktitle} {{Tau 2008,
  proceedings of the 10th International Workshop on Tau Lepton Physics,
  Novosibirsk, Russia, 22-25 September 2008}}},\ }\href {\doibase
  10.1016/j.nuclphysbps.2009.03.010} {\bibfield  {journal} {\bibinfo  {journal}
  {Nucl. Phys. Proc. Suppl.}\ }\textbf {\bibinfo {volume} {189}},\ \bibinfo
  {pages} {49} (\bibinfo {year} {2009})},\ \Eprint
  {http://arxiv.org/abs/0906.2987} {arXiv:0906.2987 [hep-ph]} \BibitemShut
  {NoStop}%
\bibitem [{\citenamefont {Chetyrkin}\ \emph {et~al.}(2000)\citenamefont
  {Chetyrkin}, \citenamefont {Harlander},\ and\ \citenamefont
  {Kuhn}}]{Chetyrkin:2000zk}%
  \BibitemOpen
  \bibfield  {author} {\bibinfo {author} {\bibfnamefont {K.~G.}\ \bibnamefont
  {Chetyrkin}}, \bibinfo {author} {\bibfnamefont {R.~V.}\ \bibnamefont
  {Harlander}}, \ and\ \bibinfo {author} {\bibfnamefont {J.~H.}\ \bibnamefont
  {Kuhn}},\ }\href {\doibase 10.1016/S0550-3213(00)00393-X,
  10.1016/S0550-3213(02)00353-X} {\bibfield  {journal} {\bibinfo  {journal}
  {Nucl. Phys.}\ }\textbf {\bibinfo {volume} {B586}},\ \bibinfo {pages} {56}
  (\bibinfo {year} {2000})},\ \bibinfo {note} {[Erratum: Nucl.
  Phys.B634,413(2002)]},\ \Eprint {http://arxiv.org/abs/hep-ph/0005139}
  {arXiv:hep-ph/0005139 [hep-ph]} \BibitemShut {NoStop}%
\bibitem [{\citenamefont {Baikov}\ \emph {et~al.}(2008)\citenamefont {Baikov},
  \citenamefont {Chetyrkin},\ and\ \citenamefont {Kuhn}}]{Baikov:2008jh}%
  \BibitemOpen
  \bibfield  {author} {\bibinfo {author} {\bibfnamefont {P.~A.}\ \bibnamefont
  {Baikov}}, \bibinfo {author} {\bibfnamefont {K.~G.}\ \bibnamefont
  {Chetyrkin}}, \ and\ \bibinfo {author} {\bibfnamefont {J.~H.}\ \bibnamefont
  {Kuhn}},\ }\href {\doibase 10.1103/PhysRevLett.101.012002} {\bibfield
  {journal} {\bibinfo  {journal} {Phys. Rev. Lett.}\ }\textbf {\bibinfo
  {volume} {101}},\ \bibinfo {pages} {012002} (\bibinfo {year} {2008})},\
  \Eprint {http://arxiv.org/abs/0801.1821} {arXiv:0801.1821 [hep-ph]}
  \BibitemShut {NoStop}%
\bibitem [{\citenamefont {Baikov}\ \emph {et~al.}(2004)\citenamefont {Baikov},
  \citenamefont {Chetyrkin},\ and\ \citenamefont {Kuhn}}]{Baikov:2004ku}%
  \BibitemOpen
  \bibfield  {author} {\bibinfo {author} {\bibfnamefont {P.~A.}\ \bibnamefont
  {Baikov}}, \bibinfo {author} {\bibfnamefont {K.~G.}\ \bibnamefont
  {Chetyrkin}}, \ and\ \bibinfo {author} {\bibfnamefont {J.~H.}\ \bibnamefont
  {Kuhn}},\ }\bibfield  {booktitle} {\emph {\bibinfo {booktitle} {{Loops and
  legs in quantum field theory. Proceedings, 7th Workshop on Elementary
  Particle Theory, Zinnowitz, Germany, April 25-30, 2004}}},\ }\href {\doibase
  10.1016/j.nuclphysbps.2004.09.013} {\bibfield  {journal} {\bibinfo  {journal}
  {Nucl. Phys. Proc. Suppl.}\ }\textbf {\bibinfo {volume} {135}},\ \bibinfo
  {pages} {243} (\bibinfo {year} {2004})},\ \bibinfo {note}
  {[,243(2004)]}\BibitemShut {NoStop}%
\bibitem [{\citenamefont {Groote}\ \emph {et~al.}(2008)\citenamefont {Groote},
  \citenamefont {Korner},\ and\ \citenamefont {Pivovarov}}]{Groote:2008dx}%
  \BibitemOpen
  \bibfield  {author} {\bibinfo {author} {\bibfnamefont {S.}~\bibnamefont
  {Groote}}, \bibinfo {author} {\bibfnamefont {J.~G.}\ \bibnamefont {Korner}},
  \ and\ \bibinfo {author} {\bibfnamefont {A.~A.}\ \bibnamefont {Pivovarov}},\
  }\href {\doibase 10.1140/epjc/s10052-008-0763-7} {\bibfield  {journal}
  {\bibinfo  {journal} {Eur. Phys. J.}\ }\textbf {\bibinfo {volume} {C58}},\
  \bibinfo {pages} {355} (\bibinfo {year} {2008})},\ \Eprint
  {http://arxiv.org/abs/0807.2148} {arXiv:0807.2148 [hep-ph]} \BibitemShut
  {NoStop}%
\bibitem [{\citenamefont {Chetyrkin}\ and\ \citenamefont
  {Tkachov}(1981)}]{Chetyrkin:1981qh}%
  \BibitemOpen
  \bibfield  {author} {\bibinfo {author} {\bibfnamefont {K.~G.}\ \bibnamefont
  {Chetyrkin}}\ and\ \bibinfo {author} {\bibfnamefont {F.~V.}\ \bibnamefont
  {Tkachov}},\ }\href {\doibase 10.1016/0550-3213(81)90199-1} {\bibfield
  {journal} {\bibinfo  {journal} {Nucl. Phys.}\ }\textbf {\bibinfo {volume}
  {B192}},\ \bibinfo {pages} {159} (\bibinfo {year} {1981})}\BibitemShut
  {NoStop}%
\bibitem [{\citenamefont {Laporta}(2000)}]{Laporta:2001dd}%
  \BibitemOpen
  \bibfield  {author} {\bibinfo {author} {\bibfnamefont {S.}~\bibnamefont
  {Laporta}},\ }\href {\doibase 10.1016/S0217-751X(00)00215-7,
  10.1142/S0217751X00002157} {\bibfield  {journal} {\bibinfo  {journal} {Int.
  J. Mod. Phys.}\ }\textbf {\bibinfo {volume} {A15}},\ \bibinfo {pages} {5087}
  (\bibinfo {year} {2000})},\ \Eprint {http://arxiv.org/abs/hep-ph/0102033}
  {arXiv:hep-ph/0102033 [hep-ph]} \BibitemShut {NoStop}%
\bibitem [{\citenamefont {Henn}(2013)}]{Henn:2013pwa}%
  \BibitemOpen
  \bibfield  {author} {\bibinfo {author} {\bibfnamefont {J.~M.}\ \bibnamefont
  {Henn}},\ }\href {\doibase 10.1103/PhysRevLett.110.251601} {\bibfield
  {journal} {\bibinfo  {journal} {Phys. Rev. Lett.}\ }\textbf {\bibinfo
  {volume} {110}},\ \bibinfo {pages} {251601} (\bibinfo {year} {2013})},\
  \Eprint {http://arxiv.org/abs/1304.1806} {arXiv:1304.1806 [hep-th]}
  \BibitemShut {NoStop}%
\bibitem [{\citenamefont {Henn}(2015)}]{Henn:2014qga}%
  \BibitemOpen
  \bibfield  {author} {\bibinfo {author} {\bibfnamefont {J.~M.}\ \bibnamefont
  {Henn}},\ }\href {\doibase 10.1088/1751-8113/48/15/153001} {\bibfield
  {journal} {\bibinfo  {journal} {J. Phys.}\ }\textbf {\bibinfo {volume}
  {A48}},\ \bibinfo {pages} {153001} (\bibinfo {year} {2015})},\ \Eprint
  {http://arxiv.org/abs/1412.2296} {arXiv:1412.2296 [hep-ph]} \BibitemShut
  {NoStop}%
\bibitem [{\citenamefont {Ioffe}(1981)}]{Ioffe:1981kw}%
  \BibitemOpen
  \bibfield  {author} {\bibinfo {author} {\bibfnamefont {B.~L.}\ \bibnamefont
  {Ioffe}},\ }\href {\doibase 10.1016/0550-3213(81)90315-1,
  10.1016/0550-3213(81)90259-5} {\bibfield  {journal} {\bibinfo  {journal}
  {Nucl. Phys.}\ }\textbf {\bibinfo {volume} {B188}},\ \bibinfo {pages} {317}
  (\bibinfo {year} {1981})},\ \bibinfo {note} {[Erratum: Nucl.
  Phys.B191,591(1981)]}\BibitemShut {NoStop}%
\bibitem [{\citenamefont {Ioffe}(1983)}]{Ioffe:1982ce}%
  \BibitemOpen
  \bibfield  {author} {\bibinfo {author} {\bibfnamefont {B.~L.}\ \bibnamefont
  {Ioffe}},\ }\href {\doibase 10.1007/BF01571709} {\bibfield  {journal}
  {\bibinfo  {journal} {Z. Phys.}\ }\textbf {\bibinfo {volume} {C18}},\
  \bibinfo {pages} {67} (\bibinfo {year} {1983})}\BibitemShut {NoStop}%
\bibitem [{\citenamefont {Chung}\ \emph {et~al.}(1982)\citenamefont {Chung},
  \citenamefont {Dosch}, \citenamefont {Kremer},\ and\ \citenamefont
  {Schall}}]{Chung:1981cc}%
  \BibitemOpen
  \bibfield  {author} {\bibinfo {author} {\bibfnamefont {Y.}~\bibnamefont
  {Chung}}, \bibinfo {author} {\bibfnamefont {H.~G.}\ \bibnamefont {Dosch}},
  \bibinfo {author} {\bibfnamefont {M.}~\bibnamefont {Kremer}}, \ and\ \bibinfo
  {author} {\bibfnamefont {D.}~\bibnamefont {Schall}},\ }\href {\doibase
  10.1016/0550-3213(82)90154-7} {\bibfield  {journal} {\bibinfo  {journal}
  {Nucl. Phys.}\ }\textbf {\bibinfo {volume} {B197}},\ \bibinfo {pages} {55}
  (\bibinfo {year} {1982})}\BibitemShut {NoStop}%
\bibitem [{\citenamefont {Leinweber}(1997)}]{Leinweber:1995fn}%
  \BibitemOpen
  \bibfield  {author} {\bibinfo {author} {\bibfnamefont {D.~B.}\ \bibnamefont
  {Leinweber}},\ }\href {\doibase 10.1006/aphy.1996.5641} {\bibfield  {journal}
  {\bibinfo  {journal} {Annals Phys.}\ }\textbf {\bibinfo {volume} {254}},\
  \bibinfo {pages} {328} (\bibinfo {year} {1997})},\ \Eprint
  {http://arxiv.org/abs/nucl-th/9510051} {arXiv:nucl-th/9510051 [nucl-th]}
  \BibitemShut {NoStop}%
\bibitem [{\citenamefont {Narison}\ and\ \citenamefont
  {Albuquerque}(2011)}]{Narison:2010py}%
  \BibitemOpen
  \bibfield  {author} {\bibinfo {author} {\bibfnamefont {S.}~\bibnamefont
  {Narison}}\ and\ \bibinfo {author} {\bibfnamefont {R.}~\bibnamefont
  {Albuquerque}},\ }\href {\doibase 10.1016/j.physletb.2010.09.051} {\bibfield
  {journal} {\bibinfo  {journal} {Phys. Lett.}\ }\textbf {\bibinfo {volume}
  {B694}},\ \bibinfo {pages} {217} (\bibinfo {year} {2011})},\ \Eprint
  {http://arxiv.org/abs/1006.2091} {arXiv:1006.2091 [hep-ph]} \BibitemShut
  {NoStop}%
\bibitem [{\citenamefont {Kublbeck}\ \emph {et~al.}(1990)\citenamefont
  {Kublbeck}, \citenamefont {Bohm},\ and\ \citenamefont
  {Denner}}]{Kublbeck:1990xc}%
  \BibitemOpen
  \bibfield  {author} {\bibinfo {author} {\bibfnamefont {J.}~\bibnamefont
  {Kublbeck}}, \bibinfo {author} {\bibfnamefont {M.}~\bibnamefont {Bohm}}, \
  and\ \bibinfo {author} {\bibfnamefont {A.}~\bibnamefont {Denner}},\ }\href
  {\doibase 10.1016/0010-4655(90)90001-H} {\bibfield  {journal} {\bibinfo
  {journal} {Comput. Phys. Commun.}\ }\textbf {\bibinfo {volume} {60}},\
  \bibinfo {pages} {165} (\bibinfo {year} {1990})}\BibitemShut {NoStop}%
\bibitem [{\citenamefont {Hahn}(2001)}]{Hahn:2000kx}%
  \BibitemOpen
  \bibfield  {author} {\bibinfo {author} {\bibfnamefont {T.}~\bibnamefont
  {Hahn}},\ }\href {\doibase 10.1016/S0010-4655(01)00290-9} {\bibfield
  {journal} {\bibinfo  {journal} {Comput. Phys. Commun.}\ }\textbf {\bibinfo
  {volume} {140}},\ \bibinfo {pages} {418} (\bibinfo {year} {2001})},\ \Eprint
  {http://arxiv.org/abs/hep-ph/0012260} {arXiv:hep-ph/0012260 [hep-ph]}
  \BibitemShut {NoStop}%
\bibitem [{\citenamefont {Mertig}\ \emph {et~al.}(1991)\citenamefont {Mertig},
  \citenamefont {Bohm},\ and\ \citenamefont {Denner}}]{Mertig:1990an}%
  \BibitemOpen
  \bibfield  {author} {\bibinfo {author} {\bibfnamefont {R.}~\bibnamefont
  {Mertig}}, \bibinfo {author} {\bibfnamefont {M.}~\bibnamefont {Bohm}}, \ and\
  \bibinfo {author} {\bibfnamefont {A.}~\bibnamefont {Denner}},\ }\href
  {\doibase 10.1016/0010-4655(91)90130-D} {\bibfield  {journal} {\bibinfo
  {journal} {Comput. Phys. Commun.}\ }\textbf {\bibinfo {volume} {64}},\
  \bibinfo {pages} {345} (\bibinfo {year} {1991})}\BibitemShut {NoStop}%
\bibitem [{\citenamefont {Shtabovenko}\ \emph {et~al.}(2016)\citenamefont
  {Shtabovenko}, \citenamefont {Mertig},\ and\ \citenamefont
  {Orellana}}]{Shtabovenko:2016sxi}%
  \BibitemOpen
  \bibfield  {author} {\bibinfo {author} {\bibfnamefont {V.}~\bibnamefont
  {Shtabovenko}}, \bibinfo {author} {\bibfnamefont {R.}~\bibnamefont {Mertig}},
  \ and\ \bibinfo {author} {\bibfnamefont {F.}~\bibnamefont {Orellana}},\
  }\href {\doibase 10.1016/j.cpc.2016.06.008} {\bibfield  {journal} {\bibinfo
  {journal} {Comput. Phys. Commun.}\ }\textbf {\bibinfo {volume} {207}},\
  \bibinfo {pages} {432} (\bibinfo {year} {2016})},\ \Eprint
  {http://arxiv.org/abs/1601.01167} {arXiv:1601.01167 [hep-ph]} \BibitemShut
  {NoStop}%
\bibitem [{\citenamefont {Smirnov}(2015)}]{Smirnov:2014hma}%
  \BibitemOpen
  \bibfield  {author} {\bibinfo {author} {\bibfnamefont {A.~V.}\ \bibnamefont
  {Smirnov}},\ }\href {\doibase 10.1016/j.cpc.2014.11.024} {\bibfield
  {journal} {\bibinfo  {journal} {Comput. Phys. Commun.}\ }\textbf {\bibinfo
  {volume} {189}},\ \bibinfo {pages} {182} (\bibinfo {year} {2015})},\ \Eprint
  {http://arxiv.org/abs/1408.2372} {arXiv:1408.2372 [hep-ph]} \BibitemShut
  {NoStop}%
\bibitem [{\citenamefont {Lee}(2014)}]{Lee:2013mka}%
  \BibitemOpen
  \bibfield  {author} {\bibinfo {author} {\bibfnamefont {R.~N.}\ \bibnamefont
  {Lee}},\ }\bibfield  {booktitle} {\emph {\bibinfo {booktitle} {{Proceedings,
  15th International Workshop on Advanced Computing and Analysis Techniques in
  Physics Research (ACAT 2013): Beijing, China, May 16-21, 2013}}},\ }\href
  {\doibase 10.1088/1742-6596/523/1/012059} {\bibfield  {journal} {\bibinfo
  {journal} {J. Phys. Conf. Ser.}\ }\textbf {\bibinfo {volume} {523}},\
  \bibinfo {pages} {012059} (\bibinfo {year} {2014})},\ \Eprint
  {http://arxiv.org/abs/1310.1145} {arXiv:1310.1145 [hep-ph]} \BibitemShut
  {NoStop}%
\bibitem [{\citenamefont {Lee}(2015)}]{Lee:2014ioa}%
  \BibitemOpen
  \bibfield  {author} {\bibinfo {author} {\bibfnamefont {R.~N.}\ \bibnamefont
  {Lee}},\ }\href {\doibase 10.1007/JHEP04(2015)108} {\bibfield  {journal}
  {\bibinfo  {journal} {JHEP}\ }\textbf {\bibinfo {volume} {04}},\ \bibinfo
  {pages} {108} (\bibinfo {year} {2015})},\ \Eprint
  {http://arxiv.org/abs/1411.0911} {arXiv:1411.0911 [hep-ph]} \BibitemShut
  {NoStop}%
\bibitem [{\citenamefont {Goncharov}(2001)}]{Goncharov:2001iea}%
  \BibitemOpen
  \bibfield  {author} {\bibinfo {author} {\bibfnamefont {A.~B.}\ \bibnamefont
  {Goncharov}},\ }\href@noop {} {\  (\bibinfo {year} {2001})},\ \Eprint
  {http://arxiv.org/abs/math/0103059} {arXiv:math/0103059 [math.AG]}
  \BibitemShut {NoStop}%
\bibitem [{\citenamefont {Peskin}(1979)}]{Peskin:1979mn}%
  \BibitemOpen
  \bibfield  {author} {\bibinfo {author} {\bibfnamefont {M.~E.}\ \bibnamefont
  {Peskin}},\ }\href {\doibase 10.1016/0370-2693(79)90129-1} {\bibfield
  {journal} {\bibinfo  {journal} {Phys. Lett.}\ }\textbf {\bibinfo {volume}
  {88B}},\ \bibinfo {pages} {128} (\bibinfo {year} {1979})}\BibitemShut
  {NoStop}%
\bibitem [{\citenamefont {Patrignani}\ \emph {et~al.}(2016)\citenamefont
  {Patrignani} \emph {et~al.}}]{Olive:2016xmw}%
  \BibitemOpen
  \bibfield  {author} {\bibinfo {author} {\bibfnamefont {C.}~\bibnamefont
  {Patrignani}} \emph {et~al.} (\bibinfo {collaboration} {Particle Data
  Group}),\ }\href {\doibase 10.1088/1674-1137/40/10/100001} {\bibfield
  {journal} {\bibinfo  {journal} {Chin. Phys.}\ }\textbf {\bibinfo {volume}
  {C40}},\ \bibinfo {pages} {100001} (\bibinfo {year} {2016})}\BibitemShut
  {NoStop}%
\bibitem [{\citenamefont {Dominguez}\ \emph {et~al.}(1994)\citenamefont
  {Dominguez}, \citenamefont {Gluckman},\ and\ \citenamefont
  {Paver}}]{Dominguez:1994ce}%
  \BibitemOpen
  \bibfield  {author} {\bibinfo {author} {\bibfnamefont {C.~A.}\ \bibnamefont
  {Dominguez}}, \bibinfo {author} {\bibfnamefont {G.~R.}\ \bibnamefont
  {Gluckman}}, \ and\ \bibinfo {author} {\bibfnamefont {N.}~\bibnamefont
  {Paver}},\ }\href {\doibase 10.1016/0370-2693(94)91027-8} {\bibfield
  {journal} {\bibinfo  {journal} {Phys. Lett.}\ }\textbf {\bibinfo {volume}
  {B333}},\ \bibinfo {pages} {184} (\bibinfo {year} {1994})},\ \Eprint
  {http://arxiv.org/abs/hep-ph/9406329} {arXiv:hep-ph/9406329 [hep-ph]}
  \BibitemShut {NoStop}%
\bibitem [{\citenamefont {Dominguez}\ \emph {et~al.}(2015)\citenamefont
  {Dominguez}, \citenamefont {Hernandez},\ and\ \citenamefont
  {Schilcher}}]{Dominguez:2014pga}%
  \BibitemOpen
  \bibfield  {author} {\bibinfo {author} {\bibfnamefont {C.~A.}\ \bibnamefont
  {Dominguez}}, \bibinfo {author} {\bibfnamefont {L.~A.}\ \bibnamefont
  {Hernandez}}, \ and\ \bibinfo {author} {\bibfnamefont {K.}~\bibnamefont
  {Schilcher}},\ }\href {\doibase 10.1007/JHEP07(2015)110} {\bibfield
  {journal} {\bibinfo  {journal} {JHEP}\ }\textbf {\bibinfo {volume} {07}},\
  \bibinfo {pages} {110} (\bibinfo {year} {2015})},\ \Eprint
  {http://arxiv.org/abs/1411.4500} {arXiv:1411.4500 [hep-ph]} \BibitemShut
  {NoStop}%
\bibitem [{\citenamefont {Aoki}\ \emph {et~al.}(2017)\citenamefont {Aoki} \emph
  {et~al.}}]{Aoki:2016frl}%
  \BibitemOpen
  \bibfield  {author} {\bibinfo {author} {\bibfnamefont {S.}~\bibnamefont
  {Aoki}} \emph {et~al.},\ }\href {\doibase 10.1140/epjc/s10052-016-4509-7}
  {\bibfield  {journal} {\bibinfo  {journal} {Eur. Phys. J.}\ }\textbf
  {\bibinfo {volume} {C77}},\ \bibinfo {pages} {112} (\bibinfo {year}
  {2017})},\ \Eprint {http://arxiv.org/abs/1607.00299} {arXiv:1607.00299
  [hep-lat]} \BibitemShut {NoStop}%
\bibitem [{\citenamefont {Albuquerque}(2013)}]{Albuquerque:2013ija}%
  \BibitemOpen
  \bibfield  {author} {\bibinfo {author} {\bibfnamefont {R.~M.}\ \bibnamefont
  {Albuquerque}},\ }\emph {\bibinfo {title} {{Charmonium Exotic States}}},\
  \href {https://inspirehep.net/record/1239350/files/arXiv:1306.4671.pdf}
  {Ph.D. thesis},\ \bibinfo  {school} {Sao Paulo U.} (\bibinfo {year} {2013}),\
  \Eprint {http://arxiv.org/abs/1306.4671} {arXiv:1306.4671 [hep-ph]}
  \BibitemShut {NoStop}%
\bibitem [{\citenamefont {Bertlmann}(1982)}]{Bertlmann:1981he}%
  \BibitemOpen
  \bibfield  {author} {\bibinfo {author} {\bibfnamefont {R.~A.}\ \bibnamefont
  {Bertlmann}},\ }\href {\doibase 10.1016/0550-3213(82)90197-3} {\bibfield
  {journal} {\bibinfo  {journal} {Nucl. Phys.}\ }\textbf {\bibinfo {volume}
  {B204}},\ \bibinfo {pages} {387} (\bibinfo {year} {1982})}\BibitemShut
  {NoStop}%
\bibitem [{\citenamefont {Sommerfeld}(1939)}]{Sommerfeld:1741612}%
  \BibitemOpen
  \bibfield  {author} {\bibinfo {author} {\bibfnamefont {A.}~\bibnamefont
  {Sommerfeld}},\ }\href {http://cds.cern.ch/record/1741612} {\emph {\bibinfo
  {title} {{Atombau und Spektrallinien}}}}\ (\bibinfo  {publisher} {Friedr.
  Vieweg \& Sohn},\ \bibinfo {address} {Braunschweig},\ \bibinfo {year}
  {1939})\BibitemShut {NoStop}%
\bibitem [{\citenamefont {Portoles}\ and\ \citenamefont
  {Ruiz-Femenia}(2002)}]{Portoles:2002rt}%
  \BibitemOpen
  \bibfield  {author} {\bibinfo {author} {\bibfnamefont {J.}~\bibnamefont
  {Portoles}}\ and\ \bibinfo {author} {\bibfnamefont {P.~D.}\ \bibnamefont
  {Ruiz-Femenia}},\ }\href {\doibase 10.1007/s10052-002-0954-6} {\bibfield
  {journal} {\bibinfo  {journal} {Eur. Phys. J.}\ }\textbf {\bibinfo {volume}
  {C24}},\ \bibinfo {pages} {439} (\bibinfo {year} {2002})},\ \Eprint
  {http://arxiv.org/abs/hep-ph/0202114} {arXiv:hep-ph/0202114 [hep-ph]}
  \BibitemShut {NoStop}%
\end{thebibliography}
\end{document}